\documentclass[11pt,a4paper]{article}
\usepackage{authblk}
\usepackage{amsmath}
\usepackage{amssymb}
\usepackage{amsthm}
\usepackage{esdiff}
\usepackage{color}
\usepackage{graphicx}
\usepackage{tikz}
\usetikzlibrary{arrows}
\usetikzlibrary{decorations}
\usepackage{natbib}
\usepackage{color}

\usepackage{epstopdf}

\title{Weight Function Approach to Study a Crack Propagating Along a Bimaterial Interface Under Arbitrary Loading in Anisotropic Solids}
\author[1,3]{L. Pryce}
\author[2]{L.Morini}
\author[1,3]{G. Mishuris}
\affil[1]{Institute of Mathematics and Physics, Aberystwyth University, Aberystwyth, Ceredigion. SY23 3BZ.}
\affil[2]{Department of Civil, Environmental and Mechanical Engineering, University of Trento, Via Mesiano 77, 38123 Trento, Italy.}
\affil[3]{Enginsoft Trento, Via della Stazione 27 - fraz. Mattarello, 38123, Trento, Italy.}
\date{\vspace{-5ex}}

\begin{document}
\maketitle
\begin{abstract}
The focus of the paper is on the study of the dynamic steady-state propagation of interfacial cracks in anisotropic bimaterials under general, non-symmetric loading conditions. Symmetric and skew-symmetric weight functions, defined as singular non-trivial solutions of a homogeneous traction-free crack problem, have been recently derived for a quasi-static semi-infinite crack at the interface between two dissimilar anisotropic materials. In this paper, the expressions for the weight functions are generalised to the case of a dynamic steady-state crack between two anisotropic media. A functional matrix equation, through which it is possible to evaluate stress intensity factors and the energy release rate at the crack tip, is obtained. A general method for calculating asymptotic coefficients of the displacement and traction fields, without any restriction regarding the loading applied on the crack faces, is developed. The proposed approach is applied for computing stress intensity factors and higher order asymptotic terms corresponding to two different examples of loading configurations acting on the crack faces in an orthotropic bimaterial.
\end{abstract}
\section{Introduction}
Evaluation of stress intensity factors and higher order asymptotic terms of displacement and stress fields represents a crucial issue for perturbative analysis of many interfacial crack problems \citep{BercialVelez, Piccolroaz10}. Recently, using a procedure based on Betti's reciprocal theorem together with weight functions \citep{Bueckner1, Bueckner2} , a general method for calculating the coefficients of the asymptotic displacements and stresses corresponding to an arbitrary loading acting on the crack faces has been developed by \citet{Piccolroaz09} for quasi-static cracks between dissimilar isotropic media, and by \citet{Morini} for interfacial cracks in two-dimensional anisotropic bimaterials. The aim of this paper is to generalize these results to the case of a dynamic steady-state crack propagation at the interface between two dissimilar anisotropic media, and to develop a general method for explicitly computing  
the coefficients in the asymptotic representations of the displacements and stresses and the energy release rate for dynamic interfacial crack problems, without any restriction regarding the loading applied at the crack faces.

The article is organized as follows: Section 2 includes some preliminary results further used in the proposed analysis. The Stroh representation of displacements and stress fields \citep{Stroh} is reported together with the Riemann-Hilbert formulation of interfacial cracks in anisotropic bimaterials developed by \citet{Suo} and \citet{Yang}. Explicit expressions for symmetric and skew-symmetric weight functions for quasi-static plane crack problems derived by \citet{Morini} and Betti's integral formula are introduced. In Section 3, weight functions matrices for a semi-infinite crack propagating at constant speed at the interface between two dissimilar orthotropic materials under plane deformation are derived. In Section 4, using explicit weight functions together with Betti integral theorem, general formulas for stress intensity factors and higher order asymptotic terms are obtained. By means of the developed approach, both symmetric and skew-symmetric loading configurations acting on the crack faces can be considered, and higher order asymptotic terms can also be computed for non-smooth loading functions. The derived stress intensity factors are then used to evaluate the energy release rate. Two illustrative examples of numerical computations for a specific asymmetric load are presented in Section 5. The effects of the loadings asymmetry on the energy release rate and the dependence of stress intensity factors on the crack tip velocity are finally discussed, and possible physical implications of these results on the continuing propagation of the crack are explored.
\section{Preliminary Results}
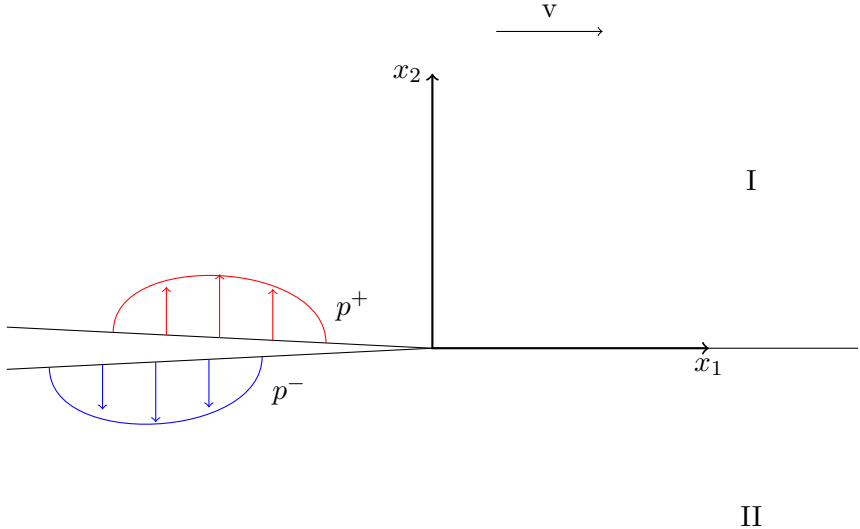
\begin{figure}
\begin{center}
\begin{tikzpicture}[scale=2.8]
\draw [-] (0,0) -- (2,0);
\draw [-] (-2,0.1) -- (0,0);
\draw [-] (-2,-0.1) -- (0,0);
\draw [->] (0.3,1.5) -- (0.8,1.5);
\node [above] at (0.55,1.52) {v};
\draw [thick, <->] (0, 1.3) -- (0,0) -- (1.3, 0);
\node [left] at (0,1.3) {$x_2$};
\node [below] at (1.3,0) {$x_1$};
\node [above] at (1.5,0.7) {I};
\node [below] at (1.5,-0.7) {II};
\draw [-,red] (-1.5,0.075) to[out=90,in=90] (-0.5,0.025);
\draw [->,red] (-1,0.05) -- (-1,0.35);
\draw [->,red] (-1.25,0.0625) -- (-1.25,0.29);
\draw [->,red] (-0.75,0.0375) -- (-0.75,0.28);
\node [right] at (-0.5,0.2) {$p^+$};
\draw [-,blue] (-1.8,-0.09) to[out=-90,in=-90] (-0.8,-0.04);
\draw [->,blue] (-1.3,-0.065) -- (-1.3,-0.35);
\draw [->,blue] (-1.55,-0.0775) -- (-1.55,-0.29);
\draw [->,blue] (-1.05,-0.0525) -- (-1.05,-0.28);
\node [right] at(-0.8,-0.2) {$p^-$};
\end{tikzpicture}
\caption{Geometry}
\label{geometry}
\end{center}
\end{figure}

In this Section the mathematical framework of the model is introduced. Preliminary results concerning interfacial cracks in two-dimensional anisotropic elastic bimaterials used for further analysis in this paper are also reported. A semi-infinite crack propagating at a constant speed,
$v$, along a perfect interface between two semi-infinite anisotropic materials is considered. The crack is said to be occupying the region $x_1-vt<0, x_2=0$ 
as illustrated in Figure \ref{geometry}.

Considering the Cartesian coordinate system shown in Figure \ref{geometry}, the traction on the crack faces is defined as follows
\begin{equation}
\sigma_{2i}(x_1 - vt, 0^{\pm})=p_j^{\pm}(x_1-vt)\quad \mbox{for} \quad x_1-vt<0,
\end{equation}
and body forces are assumed to be zero. The only restriction on the loading considered in this paper is that it must vanish within a region of the crack tip.

The closed form solution for the problem of a semi-infinite crack at an interface between two dissimilar anisotropic materials has been derived by means of Stroh formalism \citep{Stroh} both in the static \citep{Suo} and steady-state case \citep{Yang}, where the variation of angular stresses for different crack velocities was plotted. Expressions for the stress field along the interface and displacement along the crack line derived in these papers, which are used in further analysis, are reported in Section 2.1. In Section 2.2 the weight function defined in \cite{WillisMovchan} is introduced and finally, in Section 2.3, it is shown how the Betti formula can be used to relate the weight functions and the physical solutions for a problem concerning a propagating crack.

\subsection{Steady state interfacial crack: Stroh formalism}
For both anisotropic elastic media, occupying the upper and the lower half-planes in Figure \ref{geometry}, Hooke's law is given by
\begin{equation}\label{hooke}
\sigma_{ij}=C_{ijkl}\epsilon_{kl}=C_{ijkl}\frac{\partial u_k}{\partial x_l},\quad \text{for }i,j,k,l=1,2,
\end{equation}
where $\sigma$ is the stress, $\epsilon$ is the strain, $C$ is the stiffness tensor for the material, $v$ is the speed at which the crack is moving and $\rho$ is the material density.
Furthermore, the following relationship relating the stress and displacement is also used
\begin{equation}\label{stressdisp}
\sum_{j,k,l=1}^2 \frac{\partial \sigma_{ij}}{\partial x_j} = \rho\frac{\partial^2u_i}{\partial t^2}.
\end{equation} 
Combining (\ref{hooke}) and (\ref{stressdisp}) gives
\begin{equation}
C_{ijkl}\frac{\partial^2 u_k}{\partial x_j\partial x_l}=\rho\frac{\partial^2u_i}{\partial t^2}.
\end{equation}

A new coordinate system is now introduced: ($\tilde{x}_1=x_1-vt,\tilde{x}_2=x_2$). The following relationship is therefore found in this new coordinate system
\begin{equation}\label{balance}
\tilde{C}_{ijkl}\frac{\partial^2 u_k}{\partial \tilde{x}_j\partial \tilde{x}_l}=0,
\end{equation}
where $\tilde{C}_{ijkl}=C_{ijkl}-\rho v^2\delta_{ik}\delta_{1j}\delta_{1l}$.

From this stage, for convenience, the steady state coordinates will be written as $\tilde{x}_1=x$ and $\tilde{x}_2=y$. 
In order to find expressions for the displacement and stress fields in both the materials, the Stroh formalism \citep{Stroh} can be applied, and a solution in the form $u_i=A_i f(x+py)$ is derived. Introducing this expression into the balance equations (\ref{balance}), the following eigenvalue problem is obtained

\begin{equation}\label{eigenstatic}
(\mathbf{Q}+p(\mathbf{R}+\mathbf{R}^T) + p^2\mathbf{T})\mathbf{A}=0,
\end{equation}
where $\mathbf{Q}=C_{i1k1}-\rho v^2\delta_{ik}$, depends on the material constants and the crack speed. However, $\mathbf{R}=C_{i1k2}$ and $\mathbf{T}=C_{i2k2}$ depend only on elastic constants of the material. This eigenvalue problem was solved and general expressions for the traction and displacement fields can be found in \citet{Ting}.
At this stage the following matrices are also defined
\[
\mathbf{L}=(\mathbf{R}^T+p\mathbf{T})\mathbf{A},\qquad\mathbf{B}=i\mathbf{AL}^{-1},
\]
where $\mathbf{B}$ is the surface admittance tensor of the material. It is also important, for further analysis, to introduce the bimaterial matrices $\mathbf{H}$ and $\mathbf{W}$,
given by
\begin{equation}\label{hwdef}
\mathbf{H}=\mathbf{B}_I+\bar{\mathbf{B}}_{II},\qquad \mathbf{W}=\mathbf{B}_I-\bar{\mathbf{B}}_{II},
\end{equation}
where the subscript $I\text{ or }II$ determines which material the matrix relates to. It is important to note that, in the considered dynamic steady-state case, the matrices $\mathbf{A},\mathbf{L}, \mathbf{B}, \mathbf{H}$ and $\mathbf{W}$ all depend on both the elastic constants for the materials and the crack speed, $v$. 

The analysis performed in \cite{Suo} considered the static, homogeneous traction-free form of the physical problem shown in Figure \ref{geometry} with continuous traction and displacement across the interface ($x_1>0$). The work seen in \cite{Suo} has been extended to the steady-state crack by \citet{Yang} using the new coordinates $x$ and $y$ . Considering the traction-free condition, the following Riemann-Hilbert problem is satisfied along the negative portion of the real axis \citep{Suo}
\begin{equation}\label{rhnegreal}
\mathbf{h}^+(x)+\bar{\mathbf{H}}^{-1}\mathbf{H}\mathbf{h}^-(x)=0,\quad -\infty<x<0.
\end{equation}
Here, $\mathbf{h}(z)$ is a function found in the form
\[
\mathbf{h}(z)= \mathbf{w}z^{-\frac{1}{2}+i\epsilon}.
\]
The branch cut of $\mathbf{h}(z)$ is placed along the negative real axis. Combining this solution with (\ref{rhnegreal}) gives the eigenvalue problem
\begin{equation}\label{Heigen}
\bar{\mathbf{H}}\mathbf{w}=e^{2\pi\epsilon}\mathbf{H}\mathbf{w},
\end{equation}
which can be used to find $\epsilon$ and $\mathbf{w}$, both of which depend on the crack velocity \citep{Yang}.

For the positive part of the real axis the following expression for the physical traction was found in \cite{Suo}
\begin{equation}
\mathbf{t}(x)=\mathbf{h}^+(x)+\bar{\mathbf{H}}^{-1}\mathbf{H}\mathbf{h}^-(x),\quad 0<x<\infty.
\end{equation}
Combining this with the results from (\ref{Heigen}), \cite{Suo} found the following expression for the traction ahead of the crack tip
\begin{equation}\label{tracsuo}
\mathbf{t}(x)=\frac{1}{\sqrt{2\pi x}}\mathrm{Re}(Kx^{i\epsilon}\mathbf{w}),
\end{equation}
where ${K}= K_{1}+iK_{2}$ is the complex stress intensity factor, and includes both mode $I$ and mode $II$ contributions to the traction.

The displacement jump across the crack, defined as $[\mathbf{u}]=\mathbf{u}(x,0^+)-\mathbf{u}(x,0^-)$, was also found in \cite{Suo} for $x<0$
\begin{equation}\label{jumpsuo}
[\mathbf{u}](x)=\left(\frac{2(-x)}{\pi}\right)^\frac{1}{2} \frac{(\mathbf{H}+\bar{\mathbf{H}})}{\cosh\pi\epsilon} \mathrm{Re}\left(\frac{K(-x)^{i\epsilon}\mathbf{w}}{1+2i\epsilon}\right).
\end{equation}

For the physical problem with forces acting on the crack faces the asymptotic expansions of the physical traction and the jump in displacement across the interface, as $x\to0$, can be written as follows \cite{Morini}
\begin{equation}\label{asymdisjump}
[\mathbf{u}](x)=\frac{(-x)^{\frac{1}{2}}}{\sqrt{2\pi}}\boldsymbol{\mathcal{U}}(x)\mathbf{K}+\frac{(-x)^{\frac{3}{2}}}{\sqrt{2\pi}}\boldsymbol{\mathcal{U}}(x)\mathbf{Y}_2+\frac{(-x)^{\frac{5}{2}}}{\sqrt{2\pi}}\boldsymbol{\mathcal{U}}(x)\mathbf{Y}_3+\mathcal{O}((-x)^{\frac{7}{2}}),
\end{equation}
\begin{equation}\label{asymphystrac}
\mathbf{t}(x)=\frac{x^{-\frac{1}{2}}}{2\sqrt{2\pi}}\boldsymbol{\mathcal{T}}(x)\mathbf{K}+\frac{x^{\frac{1}{2}}}{2\sqrt{2\pi}}\boldsymbol{\mathcal{T}}(x)\mathbf{Y}_2+\frac{x^{\frac{3}{2}}}{2\sqrt{2\pi}}\boldsymbol{\mathcal{T}}(x)\mathbf{Y}_3+\mathcal{O}(x^{\frac{5}{2}}),
\end{equation}
where $\mathbf{K}=[K,\bar{K}]$ and $\mathbf{Y}_i=[Y_i,\bar{Y}_i]$. $Y_i$ are constants derived in the same manner as the stress intensity factor $K$ in order to find further terms in the asymptotic expansions. 
The matrices $\boldsymbol{\mathcal{U}}(x)$ and $\boldsymbol{\mathcal{T}}(x)$ are represented as follows
\begin{equation}\label{asymmat}
\boldsymbol{\mathcal{U}}(x)=\frac{2(\mathbf{H}+\bar{\mathbf{H}})}{\cosh\pi\epsilon}\left[\frac{\mathbf{w}(-x)^{i\epsilon}}{1+2i\epsilon},\frac{\bar{\mathbf{w}}(-x)^{-i\epsilon}}{1-2i\epsilon}\right],\qquad \boldsymbol{\mathcal{T}}(x)=2\left[\mathbf{w}x^{i\epsilon}, \bar{\mathbf{w}}x^{-i\epsilon}\right].
\end{equation}

An explicit formula for computing the stress intensity factor for symmetric loading was given in \cite{Suo}. It was shown that
\begin{equation}\label{sifsuo}
\mathbf{K}^S=-\left(\frac{2}{\pi}\right)^{\frac{1}{2}}\cosh \pi\epsilon\int^0_{-\infty}(-x)^{-\frac{1}{2}-i\epsilon}\langle\mathbf{p}_1\rangle(x)\mathrm{d}x,
\end{equation}
where the vector $\langle\mathbf{p}_1\rangle(x)$ is related to the applied traction $\mathbf{p}(x)$ in the following way
\[
\langle\mathbf{p}_1\rangle=\frac{\bar{\mathbf{w}}^T\mathbf{H}\langle\mathbf{p}\rangle}{\bar{\mathbf{w}}^T\mathbf{H}\mathbf{w}}.
\]
Note here that the work in \cite{Suo} only studied symmetric loading which is why the formula above only shows the part of the stress intensity factor corresponding to the symmetric part of the loading $\langle\mathbf{p}_1\rangle$. For symmetric loading the asymmetric contribution to the loading, $[\mathbf{p}_1$, is equal to 0.

Another key component in the analysis of fracture mechanics is the determination of the energy release rate (ERR) when a unit area of interface is cracked. An expression was found for the ERR, denoted $G$, in \cite{Irwin}
\begin{equation}\label{errirwin}
G=\frac{1}{2\Delta}\int_0^\Delta \mathbf{t}^T(\Delta-r)[\mathbf{u}](r)\mathrm{d}r,
\end{equation}
where $\Delta$ is an arbitrary length scale. It was stated in \cite{Yu} that this equation can still be used with an arbitrary $\Delta$ as long as the crack is moving at subsonic speeds.  It was shown in \cite{Suo}, using (\ref{tracsuo}) and (\ref{jumpsuo}), that the energy release rate can be written in the following manner
\begin{equation}\label{errsuo}
G=\frac{\bar{\mathbf{w}}^T(\mathbf{H}+\bar{\mathbf{H}})\mathbf{w}|K|^2}{4\cosh^2(\pi\epsilon)}.
\end{equation}
The value of $G$ will change as the crack moves at different speeds and this is one of the key features this paper will be studying, with the results being shown in section 5.
\subsection{Weight Functions}
The weight function $\mathbf{U}$ is now defined in the same vein as \cite{WillisMovchan}.  $\mathbf{U}=(U_1,U_2)^T$ is the singular displacement field that is obtained in the problem where the steady-state crack occupies the region of the $x$-axis with $x>0$ is now considered. Therefore $\mathbf{U}$ is discontinuous over the positive portion of the real axis. The symmetric and skew-symmetric parts of the weight function are given by
\begin{equation}
[\mathbf{U}](x)=\mathbf{U}(x,0^+)-\mathbf{U}(x,0^-),
\end{equation}\begin{equation}
\langle\mathbf{U}\rangle(x)=\frac{1}{2}(\mathbf{U}(x,0^+)+\mathbf{U}(x,0^-)).
\end{equation}
The traction field associated with the displacement field, $\mathbf{U}$, is denoted as $\mathbf{\Upsilon}= (\Upsilon_1,\Upsilon_2)^T$ and is said to be continuous over the interface ($x<0$) and the zero traction condition is imposed on the crack faces. Therefore, the following Riemann-Hilbert problem stands along the positive section of the real axis for this problem, as seen in \cite{Morini}
\begin{equation}\label{rhwfpos}
\mathbf{h}_+(x)+\bar{\mathbf{H}}^{-1}\mathbf{H}\mathbf{h}_-(x)=0,\qquad 0<x<\infty,
\end{equation}
A solution for $\mathbf{h}(z)$ is found in the form
\begin{equation}
\mathbf{h}(z)=\mathbf{v}z^{-\frac{3}{2}+i\epsilon},
\end{equation}
where the branch cut is now said to be along the positive $x$-axis. This gives the eigenvalue problem
\begin{equation}\label{Heigenopp}
\bar{\mathbf{H}}\mathbf{v}=e^{-2\pi\epsilon}\mathbf{H}\mathbf{v}.
\end{equation}
$\mathbf{H}$ is positive definite hermitian and therefore it is clear, by comparing (\ref{Heigenopp}) with (\ref{Heigen}), that $\mathbf{v}=\bar{\mathbf{w}}$. 

An expression for $\mathbf{\Upsilon}$ along the negative real axis is given by
\begin{equation}
\mathbf{\Upsilon}(x)=\mathbf{h}_+(x)+\bar{\mathbf{H}}^{-1}\mathbf{H}\mathbf{h}_-(x),\qquad -\infty<x<0.
\end{equation}
Therefore the singular traction in the steady state has the form \cite{Morini}
\begin{equation}\label{tracsingular}
\mathbf{\Upsilon}(x)=\frac{(-x)^{-\frac{3}{2}}}{{\sqrt{2\pi}}}\mathrm{Re}(R(-x)^{i\epsilon}\bar{\mathbf{w}}),
\end{equation}
where $R=R_1+iR_2$ is an arbitrary, complex number in a similar fashion to the stress intensity factor for the physical problem. 
By considering the results obtained for $\mathbf{\Upsilon}$ when $\{R_1=1,R_2=0\}$ and $\{R_1=0,R_2=1\}$ it is possible to obtain two linearly independent vectors, 
and therefore a 2x2 matrix representing $\mathbf{\Upsilon}$ \citep{Piccolroaz09}.

An expression relating the Fourier transform, defined as
\[
\hat{f}(\chi)=\int^\infty_{-\infty}f(x)e^{i\chi x}\mathrm{d}x,
\]
of the symmetric and skew-symmetric weight functions was found in \cite{Morini} following from the work seen in \cite{Piccolroaz07}
\begin{equation}\label{Usymft}
[\hat{\mathbf{U}}]^+(\chi)=\frac{1}{|\chi |}(i\mathrm{sign}(\chi)\mathrm{Im}(\mathbf{H})- \mathrm{Re}(\mathbf{H}))\hat{\Upsilon}^-(\chi),
\end{equation}
\begin{equation}\label{Uskewft}
\langle\hat{\mathbf{U}}\rangle(\chi)=\frac{1}{2|\chi |}(i\mathrm{sign}(\chi)\mathrm{Im}(\mathbf{W})- \mathrm{Re}(\mathbf{W}))\hat{\Upsilon}^-(\chi).
\end{equation}
Here the supercripts $\pm$ denotes whether the function is analytic in the upper or lower half plane respectively.

\subsection{Betti Formula}
It was mentioned previously that there are now two displacement fields to consider; the physical displacment, $\mathbf{u}$, and the singular displacement, $\mathbf{U}$. However, $\mathbf{U}$ is discontinuous across the $x$-axis for $x>0$ whereas $\mathbf{u}$ is discontinuous across the $x$-axis for $x<0$. Also considered is the traction associated with $\mathbf{U}$, given by $\mathbf{\Upsilon}$, which is continuous when $x<0$ and the traction $\mathbf{t}$ associated with $\mathbf{u}$ which is continuous when $x>0$. 

It was shown in \cite{WillisMovchan} that the Betti formula still holds for the steady state crack in isotropic materials. Using the same method it can be shown that the Betti formula still holds for the moving coordinate system in anisotropic materials. Therefore, the following expressions are found along the upper and lower parts of the real axis, respectively
\begin{equation}\label{betti+}
\int^\infty_{-\infty}{\{\mathbf{U}^T(x'-x,0^+)\boldsymbol{\mathcal{R}}\boldsymbol{\sigma}(x,0^+)-\mathbf{\Upsilon}^T(x'-x,0^+)\boldsymbol{\mathcal{R}}\mathbf{u}(x,0^+)\}\mathrm{d}x}=0,
\end{equation}
\begin{equation}\label{betti-}
\int_{-\infty}^\infty{\{\mathbf{U}^T(x'-x,0^-)\boldsymbol{\mathcal{R}}\boldsymbol{\sigma}(x,0^-)-\mathbf{\Upsilon}^T(x'-x,0^-)\boldsymbol{\mathcal{R}}\mathbf{u}(x,0^-)\}\mathrm{d}x}=0,
\end{equation}
where \[\boldsymbol{\mathcal{R}}=\begin{pmatrix}-1&0\\0&1\end{pmatrix}.\] 

The homogeneous case of (\ref{rhnegreal}) is now considered. Combined with the applied traction on the crack faces, $\mathbf{p}(x)$, the following expressions for traction are obtained
\begin{equation}\label{phystrac}
\boldsymbol{\sigma}_{2i}(x,y=0^+)=\mathbf{p}^+(x)+\mathbf{t}(x),\qquad\boldsymbol{\sigma}_{2i}(x,y=0^-)=\mathbf{p}^-(x)+\mathbf{t}(x).
\end{equation}
Subtracting (\ref{betti-}) from (\ref{betti+}) and using (\ref{phystrac}), along with the definition of the symmetric and skew-symmetric parts of the weight function, the following formula is obtained
\begin{align}
&\int_{-\infty}^\infty{\{[\mathbf{U}]^T(x'-x)\boldsymbol{\mathcal{R}}\mathbf{t}(x)-\mathbf{\Upsilon}^T(x'-x,0)\boldsymbol{\mathcal{R}}[\mathbf{u}](x)\}\mathrm{d}x}\nonumber\\
=&-\int_{-\infty}^\infty{\{[\mathbf{U}]^T(x'-x)\boldsymbol{\mathcal{R}}\langle \mathbf{p}\rangle(x)+\langle \mathbf{U}\rangle^T(x'-x)\boldsymbol{\mathcal{R}}[\mathbf{p}](x)\}\mathrm{d}x}.
\end{align}
Here, $\langle \mathbf{p}\rangle$ and $[\mathbf{p}]$ refer to the symmetric and skew-symmetric parts of the loading respectively.

Using the Fourier convolution theorem the following identity, which relates the Fourier transforms of the weight functions and the solutions of the physical problem, is obtained \cite{Piccolroaz07},\cite{Morini}
\begin{equation}\label{bettiweight}
[\hat{\mathbf{U}}]^{+T}\boldsymbol{\mathcal{R}}\hat{\mathbf{t}}^+ - \hat{\mathbf{\Upsilon}}^{-T}\boldsymbol{\mathcal{R}}[\hat{\mathbf{u}}]^-= -[\hat{\mathbf{U}}]^{+T}\boldsymbol{\mathcal{R}}\langle \hat{\mathbf{p}}\rangle-\langle\hat{\mathbf{U}}\rangle^{T}\boldsymbol{\mathcal{R}}[\hat{\mathbf{p}}],
\end{equation}
where the $\pm$ denotes whether the transform is analytic in the upper or lower half plane.%

Further work performed in \cite{Piccolroaz07} and \cite{Morini}, combining (\ref{Usymft}), (\ref{Uskewft}) and (\ref{bettiweight}), found an explicit expression for finding the stress intensity factor, $\mathbf{K}$, using the weight functions and the loading applied on the crack faces. The following expression was obtained
\begin{equation}\label{siflorenzo}
\mathbf{K}=\frac{1}{2\pi i}\boldsymbol{\mathcal{Z}}_1^{-1}\int_{-\infty}^\infty [\hat{\mathbf{U}}]^{+T}(\tau)\boldsymbol{\mathcal{R}}\langle \hat{\mathbf{p}}\rangle(\tau)+\langle\hat{\mathbf{U}}\rangle^{T}(\tau)\boldsymbol{\mathcal{R}}[\hat{\mathbf{p}}](\tau) \mathrm{d}\tau,
\end{equation}
where $\boldsymbol{\mathcal{Z}}_1$ is a constant matrix derived from the asymptotic representation of (\ref{bettiweight}). It can be shown that both expressions for $\mathbf{K}$, (\ref{sifsuo}) and (\ref{siflorenzo}), are equivalent when the loading considered is symmetric.

Following the method developed in \cite{Piccolroaz07} and \cite{Morini} an expression for further asymptotic coefficients can be found depending on whether the applied loading is smooth and has a Fourier transform that vanishes at a fast enough rate at infinity. If this is the case the general expression for the asymptotic coefficients can be found using the equation
\begin{equation}\label{secondlorenzo}
\mathbf{Y}_j=\frac{1}{2\pi i}\boldsymbol{\mathcal{Z}}_j^{-1}\int_{-\infty}^\infty \tau^{j-1}\{[\hat{\mathbf{U}}]^{+T}(\tau)\boldsymbol{\mathcal{R}}\langle \hat{\mathbf{p}}\rangle(\tau)+\langle\hat{\mathbf{U}}\rangle^{T}(\tau)\boldsymbol{\mathcal{R}}[\hat{\mathbf{p}}](\tau)\} \mathrm{d}\tau.
\end{equation}
Here, $\boldsymbol{\mathcal{Z}}_j$ is also derived from the asymptotic representation of (\ref{bettiweight}) and is found in Section 4 of this paper. 
\section{Steady-state weight functions for orthotropic bimaterials}
In this Section, expressions for the symmetric and skew-symmetric weight function matrices corresponding to steady-state plane strain interfacial crack in orthotropic bimaterials are reported. Substituting the solution for $\mathbf{w}$ found in \cite{Yang}, and shown in the Appendix of this paper, into (\ref{tracsingular}), and using the method used in \cite{Piccolroaz09}, yields the following linearly independent traction vectors for $-\infty<x<0$ 
\begin{equation}
\mathbf{\Upsilon}^1(x)= \frac{(-x)^{-\frac{3}{2}}}{2\sqrt{2\pi}}\begin{pmatrix} i[(-x)^{i\epsilon}-(-x)^{-i\epsilon}]\\ \sqrt{\frac{H_{11}}{H_{22}}}[(-x)^{i\epsilon}+(-x)^{-i\epsilon}]\end{pmatrix},\label{R=1}
\end{equation}
\begin{equation}
\mathbf{\Upsilon}^2(x)= \frac{(-x)^{-\frac{3}{2}}}{2\sqrt{2\pi}}\begin{pmatrix} -[(-x)^{i\epsilon}+(-x)^{-i\epsilon}]\\ 
i\sqrt{\frac{H_{11}}{H_{22}}}[(-x)^{i\epsilon}-(-x)^{-i\epsilon}]\end{pmatrix},\label{R=i}
\end{equation}
where $H_{11}$ and $H_{22}$ are parameters depending on the crack tip speed and elastic constants of both considered materials. Explicit expressions for $H_{11}$ and $H_{22}$ have been introduced in \cite{Yang} and are given in the Appendix. The branch cut for these vectors is situated along the positive real axis and polar coordinates with angle between $-2\pi$ and $0$ are taken. The Fourier transforms obtained are
\begin{equation}\label{ft1g}
\hat{\mathbf{\Upsilon}}^{1-}(\chi)=\frac{(i\chi)^{\frac{1}{2}}\sqrt{2}}{(1+4\epsilon^2)\sqrt{\pi}} \begin{pmatrix}
i\left[(-\frac{1}{2}-i\epsilon)\Gamma(\frac{1}{2}+i\epsilon)(i\chi)^{-i\epsilon} - (-\frac{1}{2}+i\epsilon)\Gamma(\frac{1}{2}-i\epsilon)(i\chi)^{i\epsilon}\right]\\
\sqrt{\frac{H_{11}}{H_{22}}}\left[(-\frac{1}{2}-i\epsilon)\Gamma(\frac{1}{2}+i\epsilon)(i\chi)^{-i\epsilon} + (-\frac{1}{2}+i\epsilon)\Gamma(\frac{1}{2}-i\epsilon)(i\chi)^{i\epsilon}\right]\end{pmatrix},
\end{equation}
\begin{equation}\label{ft2g}
\hat{\mathbf{\Upsilon}}^{2-}(\chi)=\frac{(i\chi)^{\frac{1}{2}}\sqrt{2}}{(1+4\epsilon^2)\sqrt{\pi}} \begin{pmatrix}
-\left[(-\frac{1}{2}-i\epsilon)\Gamma(\frac{1}{2}+i\epsilon)(i\chi)^{-i\epsilon} + (-\frac{1}{2}+i\epsilon)\Gamma(\frac{1}{2}-i\epsilon)(i\chi)^{i\epsilon}\right]\\
i\sqrt{\frac{H_{11}}{H_{22}}}\left[(-\frac{1}{2}-i\epsilon)\Gamma(\frac{1}{2}+i\epsilon)(i\chi)^{-i\epsilon} - (-\frac{1}{2}+i\epsilon)\Gamma(\frac{1}{2}-i\epsilon)(i\chi)^{i\epsilon}\right]\end{pmatrix},
\end{equation}
where $\Gamma(\cdot)$ is the gamma function and the branch cut of $\hat{\mathbf{\Upsilon}}^-$ is situated along the positive imaginary axis. Note that the expressions (\ref{ft1g}) and (\ref{ft2g}) are written using a different representation
than was used in \cite{Morini}. The reason behind this will become clearer in Section 3.

The Fourier transforms (\ref{Usymft}) and (\ref{Uskewft}) can now be computed, for $\chi\in\mathbb{R}$, with the expressions for $\mathbf{H}$ and $\mathbf{W}$ found in \cite{Yang} and \cite{Morini} respectively
\begin{equation}\label{symmweight}
[\hat{\mathbf{U}}]^+(\chi)=\frac{1}{|\chi|}\begin{pmatrix}-H_{11}&-i\beta\mathrm{sign}(\chi)\sqrt{H_{11}H_{22}}\\ 
i\beta\mathrm{sign}(\chi)\sqrt{H_{11}H_{22}}&-H_{22}\end{pmatrix}\hat{\mathbf{\Upsilon}}^-(\chi),
\end{equation}
\begin{equation}\label{skewsymmweight}
\langle\hat{\mathbf{U}}\rangle(\chi)=\frac{1}{2|\chi|}\begin{pmatrix}-\delta_1H_{11}&i\gamma\mathrm{sign}(\chi)\sqrt{H_{11}H_{22}}\\ 
-i\gamma\mathrm{sign}(\chi)\sqrt{H_{11}H_{22}}&-\delta_2H_{22}\end{pmatrix}\hat{\mathbf{\Upsilon}}^-(\chi),
\end{equation}
where branch cuts are now situated along the negative imaginary axis. Here $\beta$, $\gamma$, $\delta_1$ and $\delta_2$ are all dimensionless parameters depending on the elastic coefficients of the bimaterial and the crack tip velocity \citep{Yang}. Full expressions for both matrices, $\mathbf{H}$ and $\mathbf{W}$, are stated in the Appendix, including full expressions for the parameters $\beta$, $\gamma$, $\delta_1$ and $\delta_2$. It is clearly seen from the results of \citet{Yang} that $\beta$ is of great importance when considering the oscillations near the crack tip as $\epsilon=0$ when $\beta=0$.  
\section{Evaluation of the Coefficients in the Asymptotic Expansion of the Displacement and Stress Fields for the Steady-State Crack}
\subsection{Determination of the Stress Intensity Factor}
It is now possible to develop a method in order to find the stress intensity factor for an orthotropic bimaterial, similar to that seen for the static crack in \cite{Morini}. In the case of orthotropic materials, the matrix $\boldsymbol{\mathcal{T}}(x)$ in equation (\ref{asymphystrac}) is given by
\begin{equation}
\boldsymbol{\mathcal{T}}(x)=\begin{pmatrix}-ix^{i\epsilon}&ix^{-i\epsilon}\\
\sqrt{\frac{H_{11}}{H_{22}}}x^{i\epsilon}&\sqrt{\frac{H_{11}}{H_{22}}}x^{-i\epsilon}\end{pmatrix}.
\end{equation}
Note that this result is equivalent to (\ref{asymmat}) with the known value of $\mathbf{w}$ inserted. The Fourier transform of this expansion is computed in order to find the asymptotic expansion as $\chi\to\infty$, with $\text{Im}(\chi)\in (0,\infty)$. The result is
\begin{equation}
\hat{t}(\chi)=\frac{(-i\chi)^{-\frac{1}{2}}}{2\sqrt{2\pi}}\boldsymbol{\mathfrak{T}}_1(\chi)\mathbf{K}+\frac{(-i\chi)^{-\frac{3}{2}}}{2\sqrt{2\pi}}\boldsymbol{\mathfrak{T}}_2(\chi)\mathbf{Y}+\mathcal{O}((\chi)^{-\frac{5}{2}}),
\end{equation}
where
\begin{equation}
\boldsymbol{\mathfrak{T}}_1(\chi)=\begin{pmatrix} -i(-i\chi)^{-i\epsilon}\Gamma(\frac{1}{2}+i\epsilon)& i(-i\chi)^{i\epsilon}\Gamma(\frac{1}{2}-i\epsilon)\\
\sqrt{\frac{H_{11}}{H_{22}}}(-i\chi)^{-i\epsilon}\Gamma(\frac{1}{2}+i\epsilon) &\sqrt{\frac{H_{11}}{H_{22}}}(-i\chi)^{i\epsilon}\Gamma(\frac{1}{2}-i\epsilon)\end{pmatrix},
\end{equation}
\begin{equation}
\boldsymbol{\mathfrak{T}}_2(\chi)=\begin{pmatrix} -i(-i\chi)^{-i\epsilon}\Gamma(\frac{3}{2}+i\epsilon)& i(-i\chi)^{i\epsilon}\Gamma(\frac{3}{2}-i\epsilon)\\
\sqrt{\frac{H_{11}}{H_{22}}}(-i\chi)^{-i\epsilon}\Gamma(\frac{3}{2}+i\epsilon) &\sqrt{\frac{H_{11}}{H_{22}}}(-i\chi)^{i\epsilon}\Gamma(\frac{3}{2}-i\epsilon)\end{pmatrix}.
\end{equation}
It is noted here that these expressions differ to those seen in \cite{Morini} and \cite{Piccolroaz07} to incorporate the different branch cut used in this paper. It is now possible to find the asymptotic expansion of the members of Betti's identity from equation (\ref{bettiweight}), using expressions (\ref{symmweight}) and (\ref{skewsymmweight}), as $\chi\to\infty$
\begin{equation}\label{psi+}
[\hat{\mathbf{U}}]^{+T}\boldsymbol{\mathcal{R}}\hat{\mathbf{t}}^+ = \chi^{-1}\boldsymbol{\mathcal{Z}}_1\mathbf{K} + \chi^{-2}\boldsymbol{\mathcal{Z}}_2\mathbf{Y}_2 + \chi^{-3}\boldsymbol{\mathcal{Z}}_3\mathbf{Y}_3 + \mathcal{O}(\chi^{-4}), \qquad\text{where Im}(\chi)\in(0,\infty),
\end{equation}
\begin{equation}\label{psi-}
\hat{\mathbf{\Upsilon}}^{-T}\boldsymbol{\mathcal{R}}[\hat{\mathbf{u}}]^- = \chi^{-1}\boldsymbol{\mathcal{Z}}_1\mathbf{K} + \chi^{-2}\boldsymbol{\mathcal{Z}}_2\mathbf{Y}_2 + \chi^{-3}\boldsymbol{\mathcal{Z}}_3\mathbf{Y}_3 + \mathcal{O}(\chi^{-4}), \qquad\text{where Im}(\chi)\in(-\infty,0).
\end{equation}
The matrices $\boldsymbol{\mathcal{Z}}_1$ and $\boldsymbol{\mathcal{Z}}_2$ are given by
\[
\boldsymbol{\mathcal{Z}}_1= -\frac{H_{11}}{4s^+ s^-(1+4\epsilon^2)}
\begin{pmatrix}-\frac{(\beta - 1)(1-2i\epsilon)}{E^2} & E^2 (\beta+1)(1+2i\epsilon)\\
\frac{i(\beta - 1)(1-2i\epsilon)}{E^2}& iE^2 (\beta+1)(1+2i\epsilon)\end{pmatrix},
\]
\[
\boldsymbol{\mathcal{Z}}_2= -\frac{H_{11}}{4(1+4\epsilon^2)}
\begin{pmatrix}-\frac{(\beta - 1)(1-2i\epsilon)}{g^+ s^-E^2} & \frac{E^2 (\beta+1)(1+2i\epsilon)}{s^+ g^-}\\
\frac{i(\beta - 1)(1-2i\epsilon)}{g^+ s^-E^2}& \frac{iE^2 (\beta+1)(1+2i\epsilon)}{s^+ g^-}\end{pmatrix},
\]
where
\[ E=e^{\epsilon\frac{\pi}{2}},\qquad s^\pm=\frac{(1+i)\sqrt{\pi}}{2\Gamma\left(\frac{1}{2}\pm i\epsilon\right)},\qquad g^\pm= \frac{(1-i)\sqrt\pi}{2\Gamma\left(\frac{3}{2}\pm i\epsilon\right)}.\]

Following the method of \cite{Morini}, (\ref{bettiweight}) is rewritten as
\begin{equation}\label{psi+-}
\boldsymbol{\psi}^+(\chi)-\boldsymbol{\psi}^-(\chi)= -[\hat{\mathbf{U}}]^{+T}\boldsymbol{\mathcal{R}}\langle \hat{\mathbf{p}}\rangle-\langle\hat{\mathbf{U}}\rangle^{T}\boldsymbol{\mathcal{R}}[\hat{\mathbf{p}}],
\end{equation}
using the Plemelj formula it is possible to find $\boldsymbol{\psi}^\pm(\chi)$ using the formula
\begin{equation}
\boldsymbol{\psi}^\pm(\chi)=\frac{1}{2\pi i}\int^\infty_{-\infty} \frac{\boldsymbol{\psi}(\tau)}{\tau-\chi}\mathrm{d}\tau,
\end{equation}
where $\boldsymbol{\psi}(\tau)=-[\hat{\mathbf{U}}]^{+T}(\tau)\boldsymbol{\mathcal{R}}\langle\hat{\mathbf{p}}\rangle(\tau) -\langle\hat{\mathbf{U}}\rangle^{T}(\tau)\boldsymbol{\mathcal{R}}[\hat{\mathbf{p}}](\tau)$. The solution of (\ref{psi+-}) is given by
\[
[\hat{\mathbf{U}}]^{+T}\boldsymbol{\mathcal{R}}\hat{\mathbf{t}}^+ = \boldsymbol{\psi}^+ ,\qquad\text{where Im}(\chi)\in(0,\infty),
\]\[
\hat{\mathbf{\Upsilon}}^{-T}\boldsymbol{\mathcal{R}}[\hat{\mathbf{u}}]^- = \boldsymbol{\psi}^- ,\qquad\text{where Im}(\chi)\in(-\infty,0).
\]
The asymptotic expansion of the Plemelj formula as $\chi\to\infty^\pm$ is given by
\begin{equation}\label{plemelj}
\boldsymbol{\psi}^\pm(\chi)=\frac{1}{2\pi i}\int^\infty_{-\infty} \frac{\boldsymbol{\psi}(\tau)}{\tau-\chi}\mathrm{d}\tau= 
\chi^{-1}\mathbf{V}_1^\pm+\chi^{-2}\mathbf{V}_2^\pm+\mathcal{O}(\chi^{-3}).
\end{equation}
Comparing the terms of this asymptotic expansion with the terms of the expansions (\ref{psi+}) and (\ref{psi-}) it is clear that $\mathbf{V}_j^\pm=\boldsymbol{\mathcal{Z}}_j\mathbf{Y}_j$, where $\mathbf{Y}_1=\mathbf{K}$. 
Using (\ref{plemelj}) it is easily seen that the stress intensity factor, $\mathbf{K}$, is given by
\begin{equation}\label{Klim}
\mathbf{K}=\lim_{\chi\to\infty^\pm}\frac{1}{2\pi i}\boldsymbol{\mathcal{Z}}_1^{-1}\int_{-\infty}^\infty \frac{\chi\left(-[\hat{\mathbf{U}}]^{+T}(\tau)\boldsymbol{\mathcal{R}}\langle \hat{\mathbf{p}}\rangle(\tau)-\langle\hat{\mathbf{U}}\rangle^{T}(\tau)\boldsymbol{\mathcal{R}}[\hat{\mathbf{p}}](\tau)\right)}{\tau-\chi} \mathrm{d}\tau,
\end{equation}
where the explicit expression for $\boldsymbol{\mathcal{Z}_1}^{-1}$ is given by
\[
\boldsymbol{\mathcal{Z}_1}^{-1}=\frac{2s^+ s^- (1+4\epsilon^2)}{H_{11}}
\begin{pmatrix} \frac{E^2}{(\beta -1)(1-2i\epsilon)}&\frac{iE^2}{(\beta -1)(1-2i\epsilon)}\\
-\frac{1}{(\beta +1)(1+2i\epsilon)E^2}&\frac{i}{(\beta +1)(1+2i\epsilon)E^2}\end{pmatrix}.
\]
Assuming that the loading disappears in the region of the crack tip the limit in (\ref{Klim}) exists and therefore the general expression for the stress intensity factor, $\mathbf{K}$, for the steady state is identical to that found in \cite{Morini} (see equation (\ref{siflorenzo})).

Now that an expression for the stress intensity factor has been found it is possible to determine an the energy release rate. Using (\ref{errsuo}) the following expression is obtained for the ERR in orthotropic materials
\begin{equation}\label{Genergy}
G=\frac{H_{11}(1-\beta^2)|K|^2}{4}.
\end{equation}
\subsection{General Expression for the Coefficients of the Higher Order Terms}
Using the asymptotic expansions (\ref{psi+}), (\ref{psi-}) and the corresponding terms of and (\ref{plemelj}) a general expression for the $j$th coefficient of the asymptotic expansions, $\mathbf{Y}_i$, is found
\begin{equation}
\mathbf{V}^\pm_j=\lim_{\chi\to \infty^\pm}\left[\frac{\chi^{j}(-1)^{j-1}}{2\pi i(j-1)!}\int^\infty_{-\infty} \boldsymbol{\psi}(\tau)\frac{\mathrm{d}^{j-1}}{\mathrm{d}\chi^{j-1}}\left(\frac{\chi^{j-1}}{\tau-\chi}\right)\mathrm{d}\tau\right].
\end{equation}
This gives a general expression for the coefficients of the asymptotic expansion of the displacement and stress fields as
\begin{equation}\label{Ylim}
\mathbf{Y}_j=\lim_{\chi\to\infty^\pm}\frac{1}{2\pi i}\boldsymbol{\mathcal{Z}}_j^{-1}\int_{-\infty}^\infty \tau^{j-1}([\hat{\mathbf{U}}]^{+T}(\tau)\boldsymbol{\mathcal{R}}\langle \hat{\mathbf{p}}\rangle(\tau) +\langle\hat{\mathbf{U}}\rangle^{T}(\tau)\boldsymbol{\mathcal{R}}[\hat{\mathbf{p}}](\tau))\left(\frac{\chi}{\chi-\tau}\right)^j \mathrm{d}\tau.
\end{equation}
If the loading is applied in such a way that the limit exists it is clearly seen that equation (\ref{Ylim}) is identical to (\ref{secondlorenzo}). The limit in (\ref{Ylim}) can only be computed directly for $j\ge2$ if the loading is given by a particularly smooth function which is therefore differentiable. However, this paper considers a general loading system in which case equation (\ref{secondlorenzo}) cannot always be used. An example of loading for which (\ref{secondlorenzo}) cannot be used is when point forces are applied on the crack faces \citep{Piccolroaz09}. To find further asymptotic terms, for arbitrary loading, an alternate method must be used.

\begin{figure}
\begin{center}
\begin{tikzpicture}[scale=2.8]
\draw [->,thick] (-2,0) -- (-1,0);
\draw [->,thick] (-1,0) -- (1,0);
\draw [-,thick] (1,0) -- (2,0);
\node [above] at (-1,0.1) {$L_l$};
\node [above] at (1,0.1) {$L_r$};
\draw [->,thick] (-0.1,-2) -- (-0.05,-1);
\draw [-,thick] (-0.05,-1) -- (0,0);
\draw [-,thick] (0,0) -- (0.05,-1);
\draw [<-,thick] (0.05,-1) -- (0.1,-2);
\node [left] at (-0.1,-1.2) {$\tilde{L_l}$};
\node [right] at (0.1,-1.2) {$\tilde{L_r}$};
\draw [-] (-2,0) to[out=-90,in=180] (-0.1,-2);
\node [below] at (-1.25,-1.7) {$L_{-\infty}$};
\draw [-] (0.1,-2) to[out=0,in=-90] (2,0);
\node [below] at (1.25,-1.7) {$L_\infty$};
\draw [->] (1.5,-1.4) -- (1.9,-1.8);
\draw [->] (-1.5,-1.4) -- (-1.9,-1.8);
\draw [very thick, - ] (0,0) -- (0,-2);
\end{tikzpicture}
\caption{Integration Shift in the $\chi$-Plane}\label{intshift}
\end{center}
\end{figure}
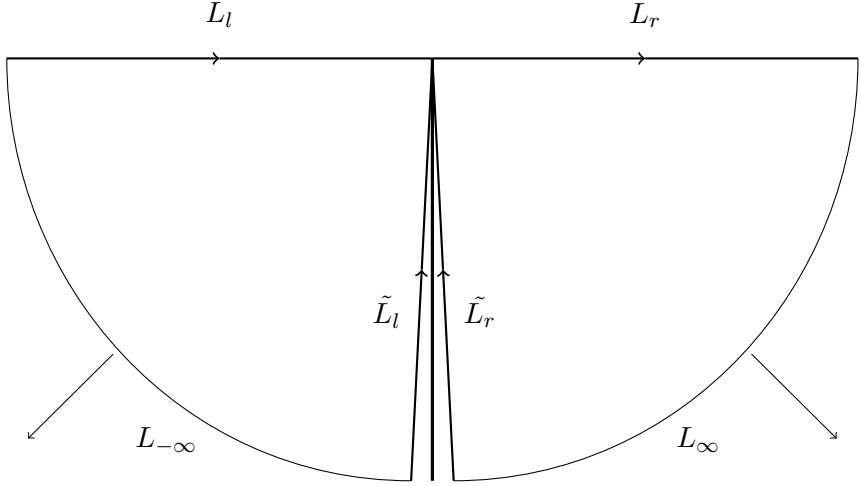 

As the function $\mathbf{p}$ only exists on the negative real $x$-axis its Fourier transform is analytic in the lower half $\chi$-plane. Therefore, $[\hat{\mathbf{p}}]$ and $\langle\hat{\mathbf{p}}\rangle$ 
are also analytic in the lower-half plane. As long as the applied loading $\mathbf{p}$ vanishes within a region of the crack tip it is clearly seen that $[\hat{\mathbf{p}}]$ and $\langle\hat{\mathbf{p}}\rangle$ 
decay exponentially as $\chi$ tends to $-i\infty$. It is also known that both $[\hat{\mathbf{U}}]^+$ and $\langle\hat{\mathbf{U}}\rangle$ are analytic in the lower-half plane apart from the negative imaginary axis. 

For computing $\mathbf{Y}_j$ the contour of integration shown in Figure \ref{intshift} is used. 
However, as there is exponential decay as $\chi$ goes to $-i\infty$, $L_{-\infty}$ and $L_\infty$ do not contribute to the total integral. Equation (\ref{Ylim}) now becomes
\begin{equation}\label{Ylimshift}
\mathbf{Y}_j=\lim_{\chi\to \infty^\pm}\left(-\frac{1}{2\pi i}\boldsymbol{\mathcal{Z}}_j^{-1}\left[\int_{\tilde{L_l}} \tau^{j-1}\boldsymbol{\psi}(\tau)\left(\frac{\chi}{\chi-\tau}\right)^j\mathrm{d}\tau - \int_{\tilde{L_r}} \tau^{j-1}\boldsymbol{\psi}(\tau)\left(\frac{\chi}{\chi-\tau}\right)^j\mathrm{d}\tau\right]\right).
\end{equation}
The limit of (\ref{Ylimshift}) can be taken to give
\begin{equation}\label{genY}
\mathbf{Y}_j=-\frac{1}{2\pi i}\boldsymbol{\mathcal{Z}}_j^{-1}\int_{-i\infty}^0 \tau^{j-1}[\boldsymbol{\psi}(\tau)]\mathrm{d}\tau,
\end{equation}
where $[\boldsymbol{\psi}(\tau)]$ refers to the jump of the function $\boldsymbol{\psi}$ over the negative imaginary axis.

The expression (\ref{genY}) can be simplified further by considering the continuity of (\ref{symmweight}) and (\ref{skewsymmweight}). 
The first term in both equations is analytic in the lower half-plane and therefore continuous over the negative imaginary axis. For this reason they do not contribute to the general
expression for the asymptotic coefficients, (\ref{genY}). Therefore, equation (\ref{genY}) simplifies to give
\begin{equation}\label{Ygensimp}
\mathbf{Y}_j=-\frac{1}{2\pi i}\boldsymbol{\mathcal{Z}}_j^{-1}\int_{-i\infty}^0 \tau^{j-1}[\boldsymbol{\phi}(\tau)]\mathrm{d}\tau,
\end{equation}
where $\boldsymbol{\phi}(\tau)$ is given by
\[
\boldsymbol{\phi}(\tau)=\frac{\mathrm{Re}(\mathbf{H})\{\hat{\Upsilon}^-(\tau)\boldsymbol{\mathcal{R}}\langle\hat{\mathbf{p}}\rangle(\tau)\}}{|\tau|}+ \frac{\mathrm{Re}(\mathbf{W})\{\hat{\Upsilon}^-(\tau)\boldsymbol{\mathcal{R}}[\hat{\mathbf{p}}](\tau)\}}{2|\tau|}.
\]
\section{Specific Examples}
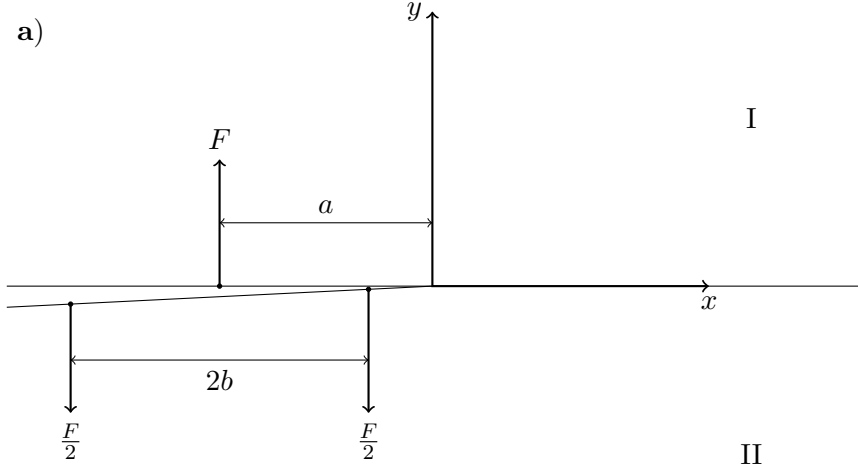
\begin{figure}
\begin{center}
\begin{tikzpicture}[scale=2.8]
\draw [-] (-2,0) -- (2,0);
\draw [-] (-2,-0.1) -- (0,0);
\draw [->,thick] (-1,0) -- (-1,0.6);
\draw [fill] (-1,0) circle [radius=0.01];
\node [above] at (-1,0.6) {$F$};
\draw [<->] (-1,0.3) -- (0,0.3);
\node [above] at (-0.5,0.3) {$a$};
\draw [->,thick] (-0.3,-0.015) -- (-0.3,-0.6);
\draw [fill] (-0.3,-0.015) circle [radius=0.01];
\node [below] at (-0.3,-0.6) {$\frac{F}{2}$};
\draw [->,thick] (-1.7,-0.085) -- (-1.7,-0.6);
\draw [fill] (-1.7,-0.085) circle [radius=0.01];
\node [below] at (-1.7,-0.6) {$\frac{F}{2}$};
\draw [<->] (-1.7,-0.35) -- (-0.3,-0.35);
\node [below] at (-1,-0.35) {$2b$};
\draw [thick, <->] (0, 1.3) -- (0,0) -- (1.3, 0);
\node [left] at (0,1.3) {$y$};
\node [below] at (1.3,0) {$x$};
\node [above] at (1.5,0.7) {I};
\node [below] at (1.5,-0.7) {II};
\node [right] at (-2,1.2) {$\mathbf{a)}$};
\end{tikzpicture}
\caption{Mode I dominant loading}\label{pointforcepic}
\end{center}
\end{figure}

Specific examples for computing the stress intensity factors for orthotropic materials are now considered. Firstly, the loading on the crack faces is given by a point force of magnitude $F$ 
acting perpendicular to the upper crack face a distance $a$ behind the crack tip and two point forces, both of magnitude $F/2$, acting perpendicular to the lower crack face a distance $b$ away from the point 
force acting upon the upper crack face. The loading moves at the same speed and in the same direction that the crack is propagating. This is shown in Figure \ref{pointforcepic}.
The forces are represented mathematically using the Dirac delta function \citep{Piccolroaz09}
\begin{equation}
p_+(x)=-F\delta(x+a),\qquad p_-(x)=-\frac{F}{2}\delta(x+a+b)-\frac{F}{2}\delta(x+a-b).
\end{equation}
It is now possible to decompose the loading into its symmetric and skew-symmetric components
\begin{align}
\langle p\rangle(x)=\frac{1}{2}[p_+(x)+p_-(x)]&=-\frac{F}{2}\delta(x+a)-\frac{F}{4}\delta(x+a-b) -\frac{F}{4}\delta(x+a-b),\nonumber\\
[p](x)=p_+(x)-p_-(x)&=-F\delta(x+a)+\frac{F}{2}\delta(x+a+b)+\frac{F}{2}\delta(x+a-b).
\end{align}
In order to compute the stress intensity factors the Fourier transforms of the skew-symmetric and symmetric parts of the loading are required. These are given by
\begin{equation}
\langle \hat{p}\rangle(\chi)=-\frac{F}{2}e^{-i\chi a}-\frac{F}{4}e^{-i\chi(a+b)}-\frac{F}{4}e^{-i\chi(a-b)},
\end{equation}
\begin{equation}
[\hat{p}](\chi)=-Fe^{-i\chi a}+\frac{F}{2}e^{-i\chi(a+b)}+\frac{F}{2}e^{-i\chi(a-b)}.
\end{equation}
It is now possible to compute expressions for the first and second order asymptotic coefficients, $\mathbf{K}$ and $\mathbf{Y}_2$, using expressions (\ref{Klim}) and (\ref{Ygensimp}) respectively.

To find an expression for $\mathbf{K}$ equation (\ref{Klim}) is used, which is identical to using the dynamic equivalent of (\ref{siflorenzo}). The solution is split into the parts corresponding to the symmetric and anti-symmetric parts of the loading, denoted $K^S$ and $K^A$ respectively
\begin{equation}\label{Kperp}
K^S_{(a)}=F\frac{E^2}{(1-\beta)}\sqrt{\frac{H_{22}}{H_{11}}}\sqrt{\frac{2}{\pi}}\,\Lambda(1,a,b,\epsilon),\quad K^A_{(a)}=F\frac{E^2\delta_2}{(1-\beta)}\sqrt{\frac{H_{22}}{H_{11}}}\sqrt{\frac{2}{\pi}}\,\Xi(1,a,b,\epsilon).
\end{equation}
where
\[
\Lambda(c,a,b,\epsilon)=a^{-\frac{c}{2}-i\epsilon}\left[\frac{1}{2}+\frac{1}{4}(1+b/a)^{-\frac{c}{2}-i\epsilon}+\frac{1}{4}(1-b/a)^{-\frac{c}{2}-i\epsilon}\right],
\]\[
\Xi(c,a,b,\epsilon)=a^{-\frac{c}{2}-i\epsilon}\left[\frac{1}{2}-\frac{1}{4}(1+b/a)^{-\frac{c}{2}-i\epsilon}-\frac{1}{4}(1-b/a)^{-\frac{c}{2}-i\epsilon}\right].
\]

Regarding higher order asymptotic coefficients for the loading shown in Figure (\ref{pointforcepic}) the alternate method developed in Section 4.2 must be used. Once again the coefficient is split into symmetric and anti-symmetric parts. The second order term is given by
\begin{equation}
Y_{2(a)}^S=F\frac{E^2}{(\beta-1)}\sqrt{\frac{H_{22}}{H_{11}}}\sqrt{\frac{2}{\pi}}\,\Lambda(3,a,b,\epsilon),\quad Y_{2(a)}^A=F\frac{E^2\delta_2}{(\beta-1)}\sqrt{\frac{H_{22}}{H_{11}}}\sqrt{\frac{2}{\pi}}\,\Xi(3,a,b,\epsilon).
\end{equation}

\begin{figure}
\begin{center}
\begin{tikzpicture}[scale=2.8]
\draw [-] (0,0) -- (2,0);
\draw [-] (-2,0.1) -- (0,0);
\draw [-] (-2,-0.1) -- (0,0);
\draw [->,thick] (-1,0.05) -- (-0.4,0.05);
\draw [fill] (-1,0.05) circle [radius=0.01];
\node [above] at (-0.1,0.05) {$F$};
\draw [<->] (-1,0.5) -- (0,0.5);
\node [above] at (-0.5,0.5) {$a$};
\draw [->,thick] (-0.3,-0.015) -- (-0.6,-0.015);
\draw [fill] (-0.3,-0.015) circle [radius=0.01];
\node [below] at (-0.3,-0.015) {$\frac{F}{2}$};
\draw [->,thick] (-1.7,-0.085) -- (-2,-0.085);
\draw [fill] (-1.7,-0.085) circle [radius=0.01];
\node [below] at (-1.7,-0.085) {$\frac{F}{2}$};
\draw [<->] (-1.7,-0.35) -- (-0.3,-0.35);
\node [below] at (-1,-0.35) {$2b$};
\draw [thick, <->] (0, 1.3) -- (0,0) -- (1.3, 0);
\node [left] at (0,1.3) {$y$};
\node [below] at (1.3,0) {$x$};
\node [above] at (1.5,0.7) {I};
\node [below] at (1.5,-0.7) {II};
\node [right] at (-2,1.2) {$\mathbf{b)}$};
\end{tikzpicture}
\caption{Mode II dominant loading}\label{parallelforcepic}
\end{center}
\end{figure}
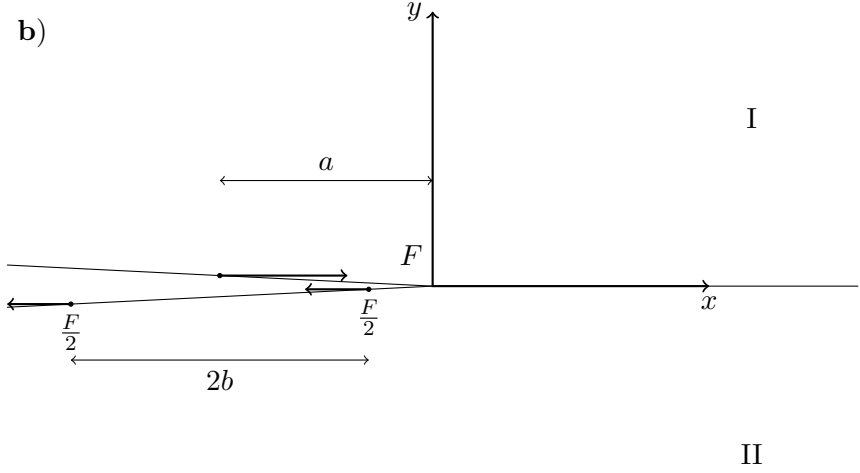

A different configuration has also been considered. This other point loading system consists of point forces acting on the crack faces at the same points as previously considered but the forces are now running parallel to the crack as opposed to the perpendicular system shown in Figure \ref{pointforcepic}. This different loading is shown in Figure \ref{parallelforcepic}.

For this loading the following expressions are found for the symmetric and antisymmetric part of the stress intensity factors
\begin{equation}\label{Kparallel}
K^S_{(b)}=iF\frac{E^2}{(1-\beta)}\sqrt{\frac{2}{\pi}}\,\Lambda(1,a,b,\epsilon),\quad K^A_{(b)}=iF\frac{E^2\delta_1}{(1-\beta)}\sqrt{\frac{2}{\pi}}\,\Xi(1,a,b,\epsilon).
\end{equation}
Using the method developed in Section 4.2, the symmetric and antisymmetric components of the second order asymptotic coefficient are found
\begin{equation}
Y_{2(b)}^S=iF\frac{E^2}{(\beta-1)}\sqrt{\frac{2}{\pi}}\,\Lambda(3,a,b,\epsilon),\quad Y_{2(b)}^A=iF\frac{E^2\delta_1}{(\beta-1)}\sqrt{\frac{2}{\pi}}\,\Xi(3,a,b,\epsilon).
\end{equation}

Having computed expressions for the stress intensity factors it is now possible to calculate the energy release rate for two given materials.  The velocity is normalised by dividing by $c_R$, the lowest of the two Rayleigh wave speeds for the given materials. This is done because the Rayleigh wave speed is a limiting velocity for which the steady-state coordinate system can be used. In the results shown the energy release rate is normalised in the following manner: $GC_{66}^{(1)}/F^2$. Here, $C_{66}^{(1)}$ is taken as the value of $C_{66}$ for the material above the crack. In all figures in this paper graphs labelled $\mathbf{a)}$ correspond to the mode I dominant loading whereas those labelled $\mathbf{b)}$ refer to the case with mode II dominant loading. For the purpose of calculations, $a$ is set as $1$ in this paper.

For this paper material I is the piezoceramic Barium Titanate. Information on this material has been obtained from \citet{Geis} which states that the material is transverse isotropic, which is a subgroup of orthotropic materials. Material II is set as Aluminium, with a cubic structure, where material paramaters have been obtained from \citet{Bower}. The properties of these materials are shown in Table \ref{matprop}. Using the method outlined in the Appendix it can be shown that the Rayleigh wave speed of Barium Titanate is $1,771$ ms$^{-1}$ and for Aluminium it is $2,941$ ms$^{-1}$. Therefore the normalising velocity, $c_R$, used is that of Barium Titanate.
\begin{table}[ht]
\begin{center}
\begin{tabular}{| l | c | c | c | c | c |}
\hline
Material & $C_{11}$(GPa) & $C_{22}$(GPa) & $C_{12}$(GPa) & $C_{66}$(GPa) & $\rho$(kgm$^{-3})$\\ \hline
I. Barium Titanate & 120.3 & 120.3 & 75.2 & 21.0 & 6,020\\ \hline
II. Aluminium & 107.3 & 107.3 & 60.9 & 28.3 & 2,700\\ \hline
\end{tabular}
\caption{Material properties}
\label{matprop}
\end{center}
\end{table}

\begin{figure}[ht]
\begin{center}
\begin{minipage}{0.9\linewidth}
	\hspace{10mm}
	\includegraphics[width=0.85\linewidth]{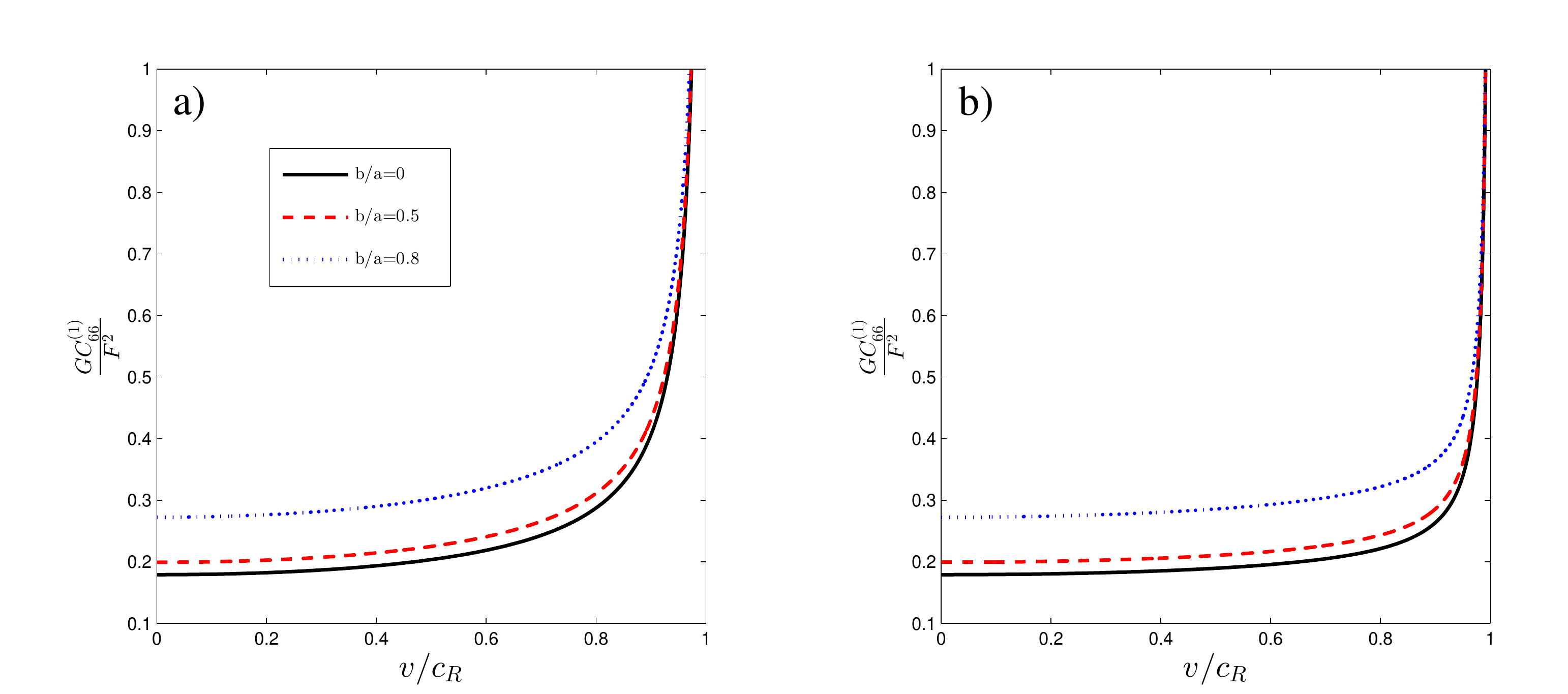}
\end{minipage}
\caption{The normalised ERR, as a function of the velocity, for different positions of the self-balanced point forces applied to the crack surfaces, described by the ratio $b/a$.}\label{energypic}
\end{center}
\end{figure}
\begin{figure}[ht]
\begin{center}
\begin{minipage}{0.9\linewidth}
	\hspace{10mm}
	\includegraphics[width=0.85\linewidth]{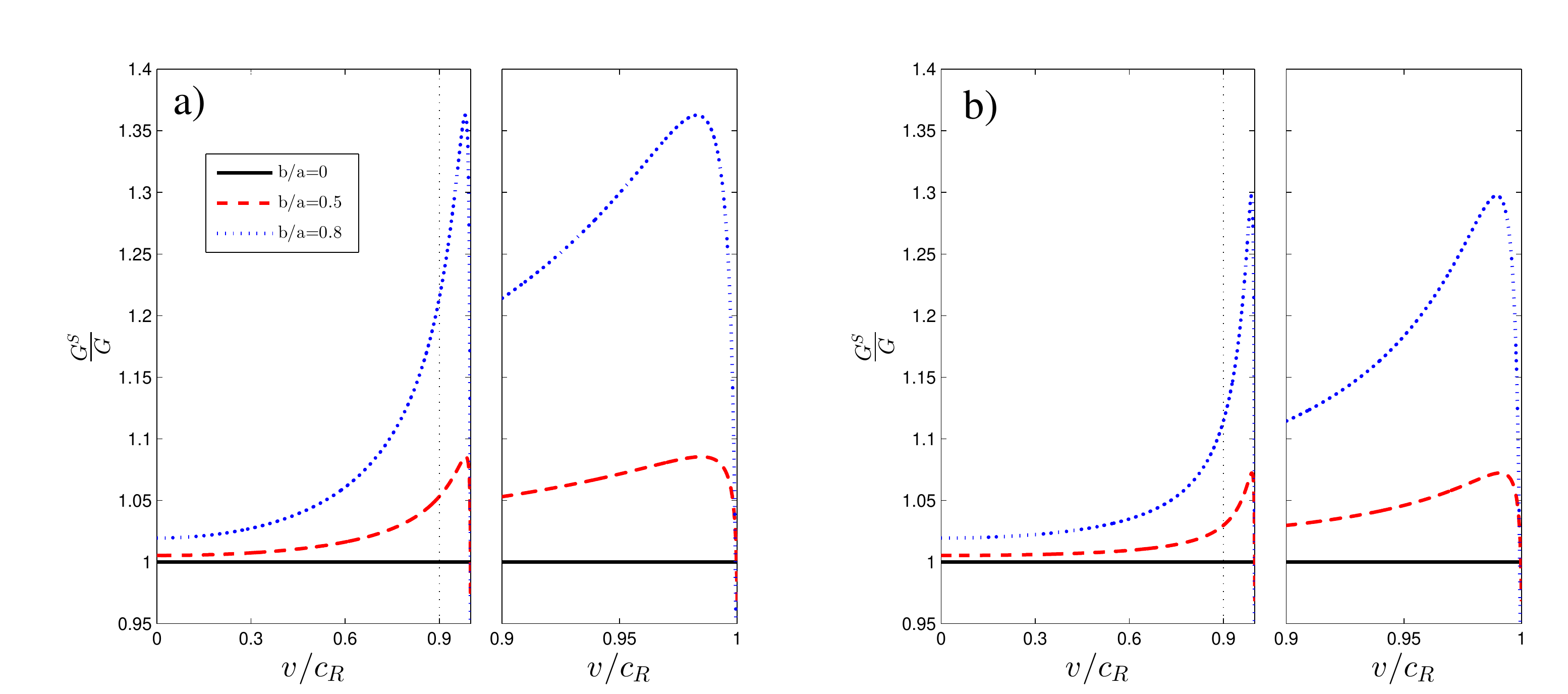}
\end{minipage}
\caption{The normalised symmetric part of the ERR, as a function of the velocity, for different positions of the self-balanced point forces applied to the crack surfaces, described by the ratio $b/a$.}\label{symmetricratio}
\end{center}
\end{figure}
\begin{figure}[ht]
\begin{center}
\begin{minipage}{0.9\linewidth}
	\hspace{10mm}
	\includegraphics[width=0.85\linewidth]{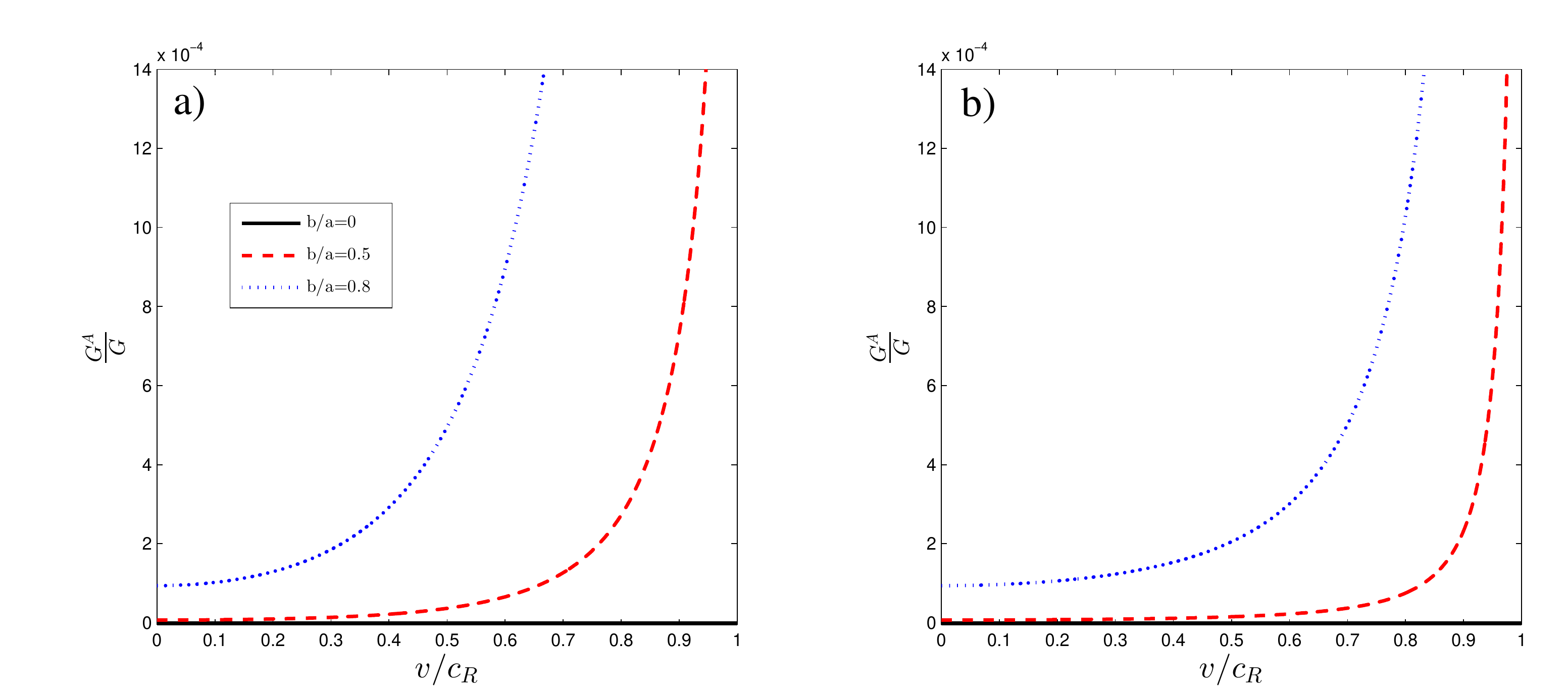}
\end{minipage}
\caption{The normalised antisymmetric part of the ERR, as a function of the velocity, for different positions of the self-balanced point forces applied to the crack surfaces, described by the ratio $b/a$.}\label{antisymmetricratio}
\end{center}
\end{figure}
\begin{figure}[ht]
\begin{center}
\begin{minipage}{0.9\linewidth}
	\hspace{10mm}
	\includegraphics[width=0.85\linewidth]{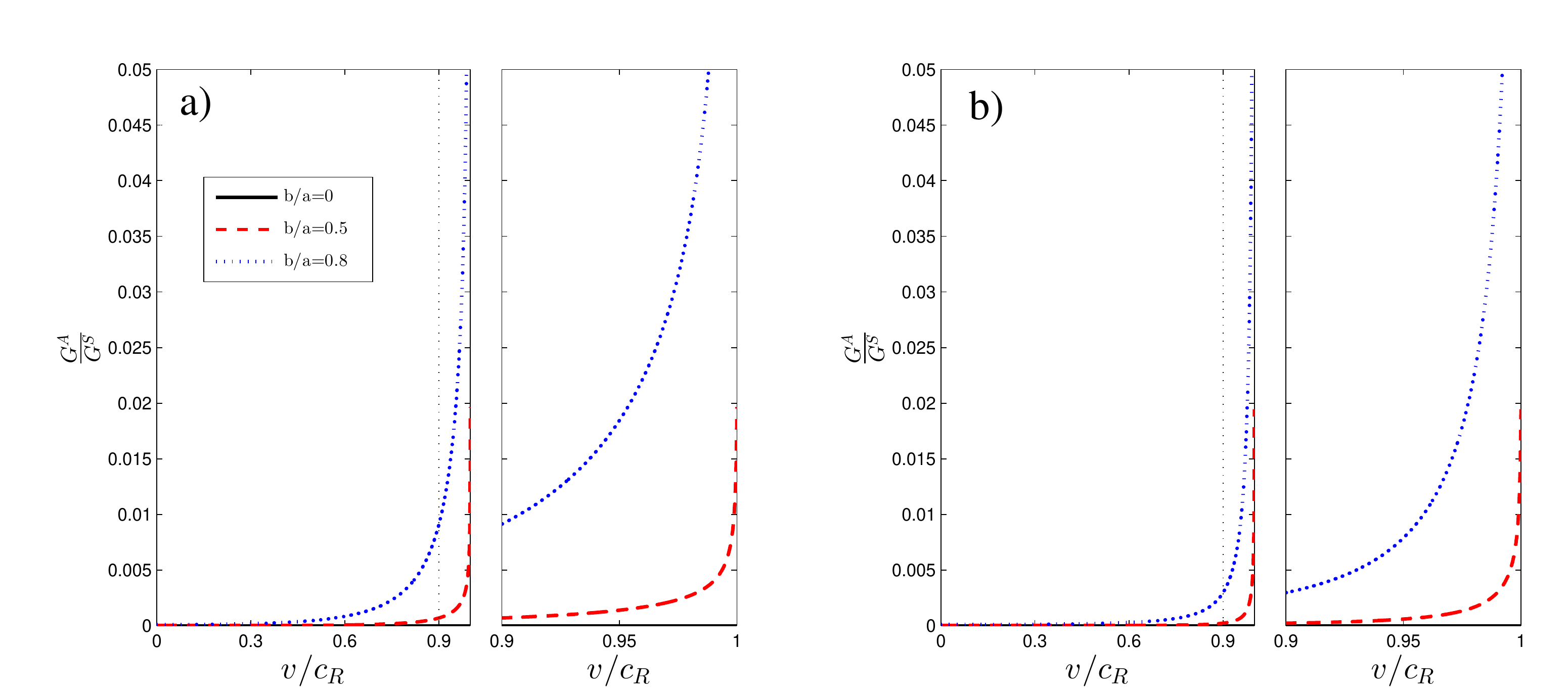}
\end{minipage}
\caption{The ratio of antisymmetric and symmetric parts of the energy release rate, as a function of the velocity, for different positions of the self-balanced point forces applied to the crack surfaces, described by the ratio $b/a$.}\label{symmoverant}
\end{center}
\end{figure}
\begin{figure}[ht]
\begin{center}
\begin{minipage}{0.95\linewidth}
	\hspace{10mm}
	\includegraphics[width=0.9\linewidth]{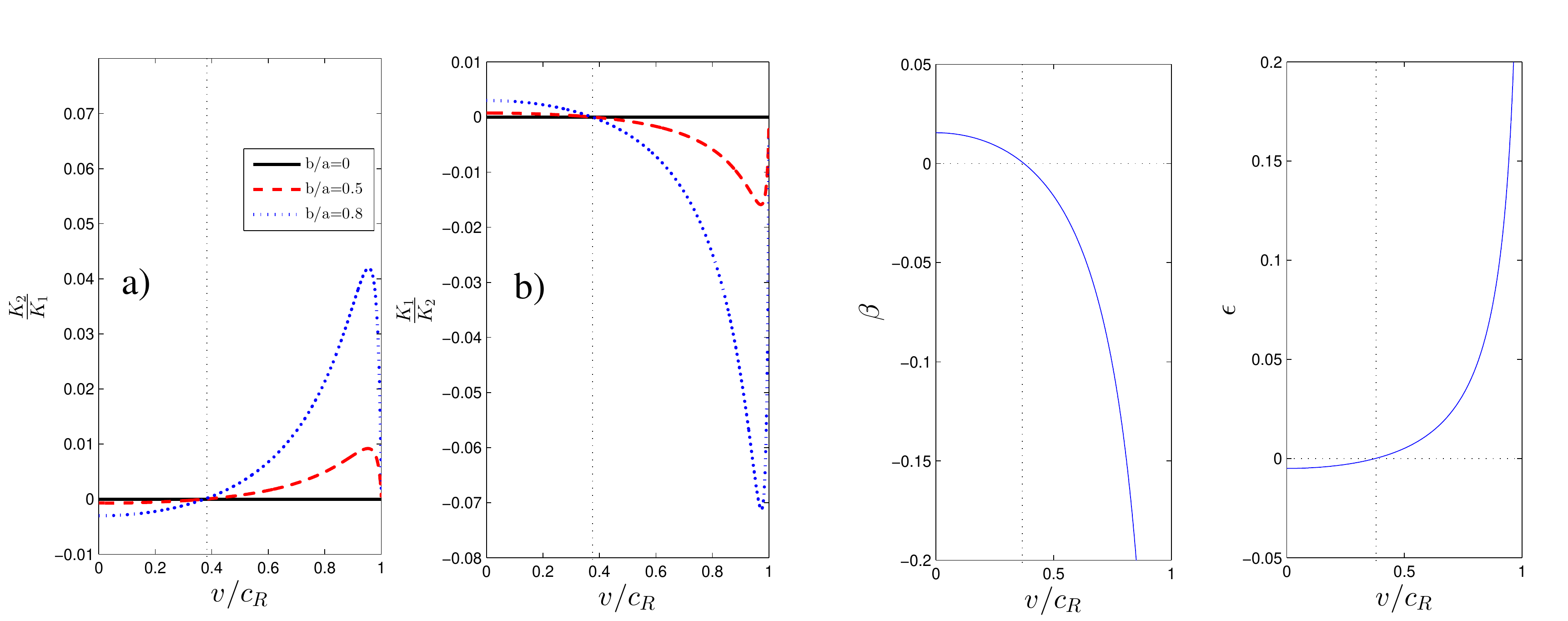}
\end{minipage}
\caption{The ratios $K_2/K_1$ and $K_1/K_2$ for the mode I and mode II loadings respectively. The graphs of $\beta$ and $\epsilon$, as a function of velocity, are also shown.}\label{sifbeta}
\end{center}
\end{figure}

\begin{figure}[ht]
\begin{center}
\begin{minipage}{0.9\linewidth}
	\hspace{10mm}
	\includegraphics[width=0.85\linewidth]{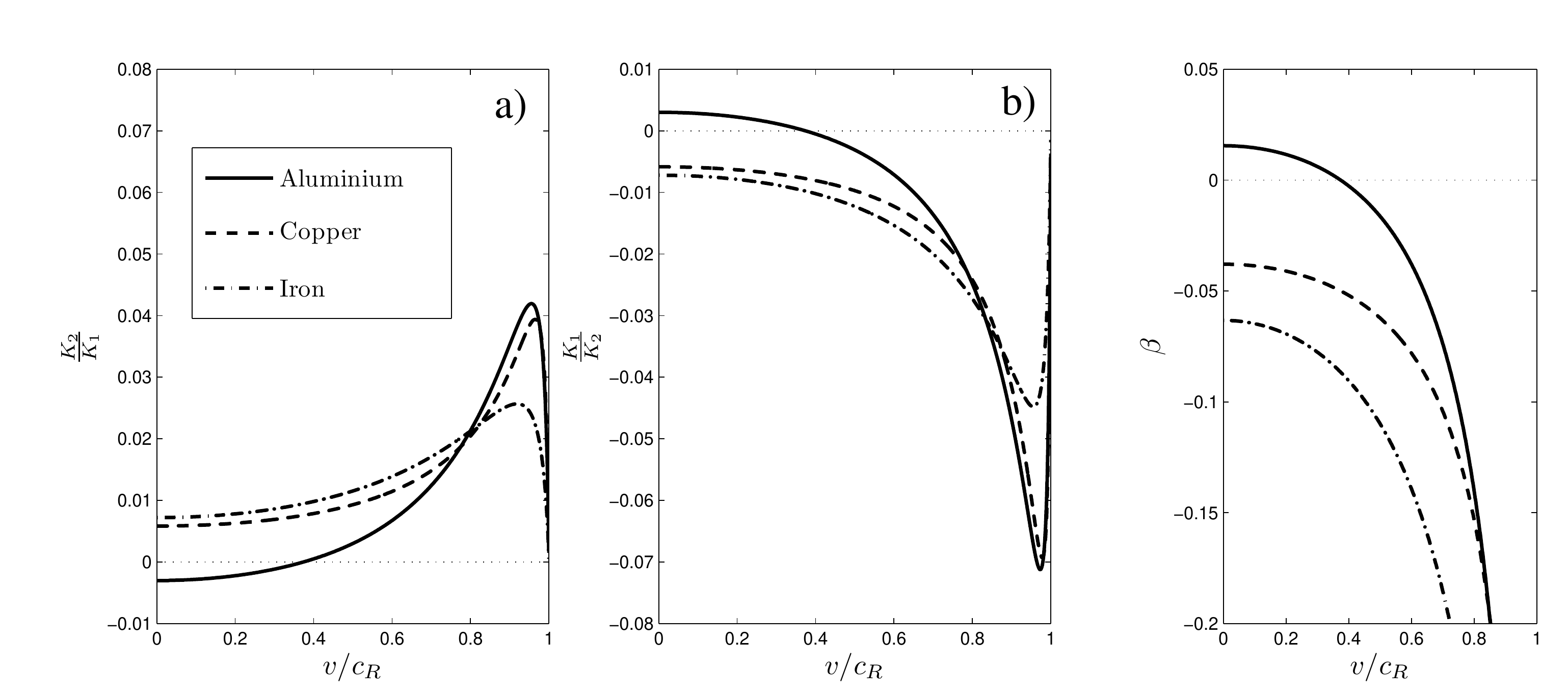}
\end{minipage}
\caption{The change in behaviour of the crack propagation when the material below the crack is changed, for fixed asymmetry of the loading.}\label{diffmats}
\end{center}
\end{figure}

\begin{figure}[ht]
\begin{center}
\begin{minipage}{0.9\linewidth}
	\hspace{10mm}
	\includegraphics[width=0.85\linewidth]{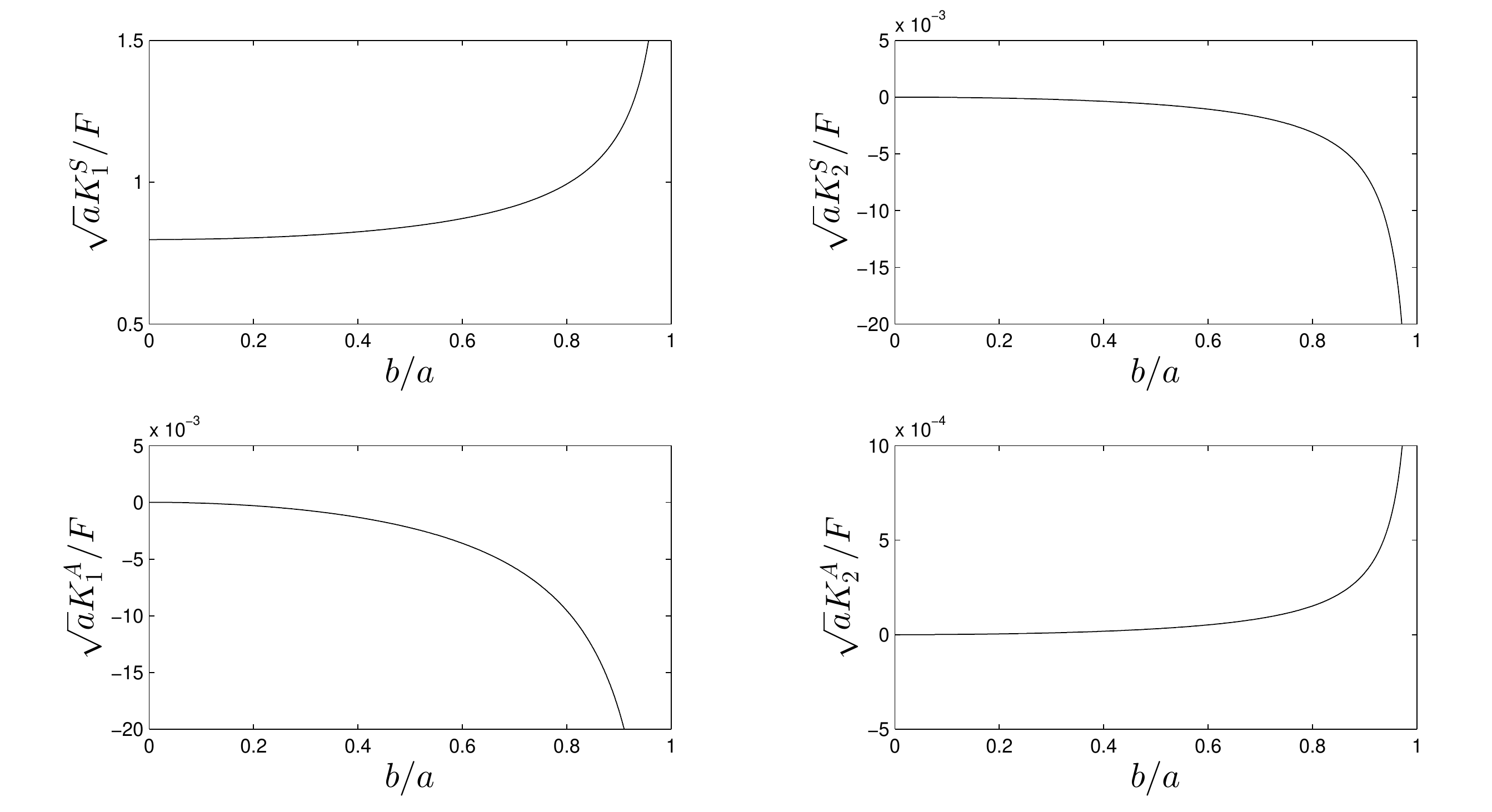}
\end{minipage}
\caption{The normalised components of $K^S$ and $K^A$ for $v=0$ with mode I dominant loading.}\label{0velocity}
\end{center}
\end{figure}

Figure \ref{energypic} shows the variation of the normalised energy release rate, as a function of the velocity, for both loadings considered, whereas Figures \ref{symmetricratio} and \ref{antisymmetricratio} illustrate the symmetric and antisymmetric contribution to the ERR, corresponding to $K^S$ and $K^A$ respectively. Both $G^S$ and $G^A$ are normalised by the total energy release rate $G$, associated with $K=K^S+K^A$.

It can be observed in Figure \ref{energypic} that the energy release rate increases as the velocity increases and tends towards infinity as the velocity approaches the Rayleigh wave speed. This behaviour is observed regardless of the asymmetry of the loading acting on the crack faces. It is important to note that, as velocity increases, asymmetry gives a larger ERR, therefore it can be said that symmetric loading 
is more energetically beneficial than any asymmetric load.

Graphs in Figures \ref{symmetricratio} and \ref{antisymmetricratio} show that for $b/a=0$, when both loadings become symmetric, $G^S/G=1$ and $G^A/G=0$ therefore the energy release rate only consists of its symmetric part, regardless of velocity, which agrees with the results found for isotropic and anisotropic
bimaterials in \cite{Piccolroaz09} and \citet{Morini}. When asymmetry is introduced into the loading it is observed that the symmetric contribution to the energy release rate is higher than the total ERR and the ratio increases 
as the velocity increases. Upon approaching the Rayleigh wave speed there is an unexpected sharp decrease in the ratio $G^S/G$. This effect should be studied further by performing experiments studying crack propagation at near-Rayleigh speeds.

In comparison to the symmetric contribution shown in Figure \ref{symmetricratio}, the asymmetric part of the ERR, illustrated in Figure \ref{antisymmetricratio}, is very small, in particular for low velocities. As the velocity starts to increase the asymmetric contribution to $G$ becomes larger. This result is supported by Figure \ref{symmoverant}, showing the ratio $G^A/G^S$, which also shows an increased contribution by the asymmetric part of the loading at higher velocities.

The dependence of the stress intensity factor, $K$, on the normalised crack tip speed is illustrated in Figure \ref{sifbeta}. The first graph shows the ratio $K_2/K_1$ for the mode 1 dominant loading. Here, $K_1$ and $K_2$ are the mode 1 and 2 contributions to the SIF, respectively.
For symmetric loading there is no mode 2 contribution to $K$, due to the fact that there is only mode  1 opening of the crack. It is important to observe that if asymmetry is introduced, for all values of $b/a$, there exists a velocity at 
which $K_2$ changes sign. The second image in Figure \ref{sifbeta} shows a similar result for the mode 2 dominant loading considering the ratio $K_1/K_2$. In this case, it is the $K_1$ component which changes sign. 
The velocity at which this change takes place is the same for both types of loading and does not depend on the asymmetry. This velocity corresponds to the value of the crack tip speed at which the Dundurs parameter, $\beta$, vanishes. This characteristic velocity can be found by solving the algebraic equation $\beta(v)=0$ and depends only on the elastic properties of the materials and the speed at which the crack is propagating while the asymmetry of the load does not affect the value at which the stress intensity factors have a change in sign. It is also clear from equation \eqref{eigensol}, found in the appendix of this paper, that when $\beta$ vanishes the oscillatory term, $\epsilon$, vanishes and this has also been shown in Figure \ref{sifbeta}. This agrees with the obtained results as, when $\epsilon=0$, it can be observed that \eqref{Kperp} consists only of real terms and \eqref{Kparallel} only has imaginary components.

It can be said that, when the crack tip speed reaches this characteristic value of the velocity, associated with $\beta=0$, the propagation should continue along the interface in a straight line. Instead, when neither $K_1$ or $K_2$ are 0 there is a possibility of kinking or
branching of the propagation. Increased magnitudes of the ratios considered in Figure \ref{sifbeta} lead to an increased probability of crack redirection and as the velocity increases the ratios exhibit this behaviour which 
explains why straight propagation along the interface is unlikely for high crack speeds. These results are in agreement with many theoretical and experimental studies which have demonstrated that there exists a specific sub-Rayleigh 
velocity which is related to the stability of the crack propagation \citep{Obrezanova2,Obrezanova1}. 

The behaviour of the stress intensity factor is also observed in Figure \ref{diffmats} for different materials in the lower half plane. The asymmetry of the load was fixed at $b/a=0.8$. The results in these graphs show that the previously mentioned speed at which the direction of the crack propagation changes does not exist for all bimaterials. This is due to the fact that there does not always exist a velocity at which $\beta=0$. For bimaterials which do not have this characteristic velocity the change of behaviour of the crack propagation would not be expected. However, the increased probability of kinking/branching at higher velocities is still observed.

Figure \ref{0velocity} shows the variation in the real and imaginary parts of the normalised stress intensity factor when $v=0$ and the asymmetry of the loading is varied. The loading considered here is the mode I dominant loading so a comparison can be made to the results obtained for this system in \citet{Morini}. The results shown agree with those in \citet{Morini} with only the real part of the symmetric stress intensity factor existing for symmetric loading and the magnitude of all components increases as the asymmetry becomes more profound. The behaviour is not identical to that seen in \citet{Morini} due to the different materials considered in this paper.

\section{Conclusions}
A general method for calculating stress intensity factors and higher order terms in the asymptotic expansions of the displacement and stress fields for a dynamic steady-state crack at the interface between two dissimilar anisotropic materials has been developed. The proposed approach, based on weight functions theory and Betti integral formula, can be applied to many crack problems in a wide range of materials, for example, several classes of anisotropic elastic media (monoclinic, orthotropic) and piezoceramics. As a particular case,  a steady-state plane interfacial crack  in orthotropic bimaterials has been studied. Expressions for the SIF and further higher order asymptotic coefficients have been found for two different configurations of loading acting on the crack faces. 

It has been shown in the considered examples that greater asymmetry of the loading configuration leads to an increase in the energy release rate at the crack tip and has a particularly large effect for high crack velocities. Moreover, the analysis of the stress intensity factors for both loadings shows the existence of a sub-Rayleigh velocity at which the SIF changes sign which could lead to a change in direction in the crack propagation. This effect is only observable when asymmetric loading was applied and may give some explanation to the fact that kinking/branching is more probable at certain velocities. As different materials for the lower half-plane are considered, it has been shown that this characteristic velocity does not exist for every bimaterial and therefore experimental study is of great importance in order to clearly detect the presence of this critical value and its physical implications on crack propagation stability.

\section*{Acknowledgments}
LP, and GM acknowledge support from the FP7 IAPP project `INTERCER2', project reference PIAP-GA-2011-286110-INTERCER2.  LM gratefully thanks financial support from the Italian Ministry of Education, University and Research in the framework of the FIRB project 2010 "Structural mechanics models for renewable energy applications". The authors would also like to acknowledge the pleasant work environment provided at Enginsoft, Trento.
\cleardoublepage

\bibliography{sources}

\begin{thebibliography}{18}
\expandafter\ifx\csname natexlab\endcsname\relax\def\natexlab#1{#1}\fi
\expandafter\ifx\csname url\endcsname\relax
  \def\url#1{\texttt{#1}}\fi
\expandafter\ifx\csname urlprefix\endcsname\relax\def\urlprefix{URL }\fi

\bibitem[{Bercial-Velez et~al.(2005)Bercial-Velez, Antipov, and
  Movchan}]{BercialVelez}
Bercial-Velez, J.~P., Antipov, Y.~A., Movchan, A.~B., 2005. High-order
  asymptotics and perturbation problems for 3d interfacial cracks. J. Mech.
  Phys. Solids 53, 1128--1162.

\bibitem[{Bower(2009)}]{Bower}
Bower, A.~F., 2009. Applied mechanics of solids, 1st Edition. CRC Press, Boca
  Raton, Florida.

\bibitem[{Bueckner(1985)}]{Bueckner1}
Bueckner, H.~F., 1985. Weight functions and fundamental fields for the
  penny-shaped and the half plane crack in three-space. Int. J. Solids Struct.
  23, 57--93.

\bibitem[{Bueckner(1989)}]{Bueckner2}
Bueckner, H.~F., 1989. Observations on weight functions. Eng. Anal. Bound.
  Elem. 6, 3--18.

\bibitem[{Geis et~al.(2004)Geis, Mishuris, and Sandig}]{Geis}
Geis, W., Mishuris, G., Sandig, A., 2004. Asymptotic models for piezoelectric
  stack actuators with thin metal inclusions. Preprint 2004/001, Univeristy of
  Stuttgart, http://preprints.ians.uni-stuttgart.de.

\bibitem[{Irwin(1957)}]{Irwin}
Irwin, G.~R., 1957. Analysis of stresses and strains near the end of a crack
  traversing a plate. J. Appl. Mech 24, 361--364.

\bibitem[{Morini et~al.(2013)Morini, Radi, Movchan, and Movchan}]{Morini}
Morini, L., Radi, E., Movchan, A.~B., Movchan, N.~V., 2013. Stroh formalism in
  analysis of skew-symmetric and symmetric weight functions for interfacial
  cracks. Math. Mech. Solids 18, 135--152.

\bibitem[{Obrezanova et~al.(2002{\natexlab{a}})Obrezanova, Willis, and
  Movchan}]{Obrezanova2}
Obrezanova, O., Willis, J.~R., Movchan, A.~B., 2002{\natexlab{a}}. Dynamic
  stability of a propagating crack. J. Mech. Phys. Solids 50, 2637--2668.

\bibitem[{Obrezanova et~al.(2002{\natexlab{b}})Obrezanova, Willis, and
  Movchan}]{Obrezanova1}
Obrezanova, O., Willis, J.~R., Movchan, A.~B., 2002{\natexlab{b}}. Stability of
  an advanicng crack to small perturbation of its path. J. Mech. Phys. Solids
  50, 57--80.

\bibitem[{Piccolroaz et~al.(2007)Piccolroaz, Mishuris, and
  Movchan}]{Piccolroaz07}
Piccolroaz, A., Mishuris, G., Movchan, A.~B., 2007. Evaluation of the
  lazarus-leblond constants in the asymptotic model for the interfacial wavy
  crack. J. Mech. Phys. Solids 55, 1575--1600.

\bibitem[{Piccolroaz et~al.(2009)Piccolroaz, Mishuris, and
  Movchan}]{Piccolroaz09}
Piccolroaz, A., Mishuris, G., Movchan, A.~B., 2009. Symmetric and
  skew-symmetric weight functions in 2d perturbation models for semi-infinite
  interfacial cracks. J. Mech. Phys. Solids 57, 1657--1682.

\bibitem[{Piccolroaz et~al.(2010)Piccolroaz, Mishuris, and
  Movchan}]{Piccolroaz10}
Piccolroaz, A., Mishuris, G., Movchan, A.~B., 2010. Perturbation of mode iii
  interfacial cracks. Int. J. Fract. 166, 41--51.

\bibitem[{Stroh(1962)}]{Stroh}
Stroh, A.~N., 1962. Steady state problems in anisotropic elasticity. Math. Phys
  41, 77--103.

\bibitem[{Suo(1990)}]{Suo}
Suo, Z., 1990. Singularities, interfaces and cracks in dissimilar anisotropic
  media. Proc. R. Soc. Lond 427, 331--358.

\bibitem[{Ting(1996)}]{Ting}
Ting, T. C.~T., 1996. Anisotropic elasticity: theory and applications. Oxford
  University Press.

\bibitem[{Willis and Movchan(1995)}]{WillisMovchan}
Willis, J.~R., Movchan, A.~B., 1995. Dynamic weight function for a moving
  crack. i. mode i loading. J. Mech. Phys. Solids, 319--341.

\bibitem[{Yang et~al.(1991)Yang, Suo, and Shih}]{Yang}
Yang, W., Suo, Z., Shih, C.~F., 1991. Mechanics of dynamic debonding. Proc.
  Mathematical and Physical Sciences 433, 679--697.

\bibitem[{Yu and Suo(2000)}]{Yu}
Yu, H.~H., Suo, Z., 2000. Intersonic crack growth on an interface. Proc. R.
  Soc. Lond 456, 223--246.

\end{thebibliography}
\bibliographystyle{elsarticle-harv}

\appendix
\section{Orthotropic Stroh matrices for a dynamic crack}
For orthotropic materials the matrices $\mathbf{Q},\mathbf{R}$ and $\mathbf{T}$ are given by
\begin{equation}\label{orthmat}\mathbf{Q}=\begin{pmatrix}C_{11}-\rho v^2&0\\0&C_{66}-\rho v^2\end{pmatrix}, \mathbf{R}=\begin{pmatrix}0&C_{12}\\C_{66}&0\end{pmatrix},
\mathbf{T}=\begin{pmatrix}C_{66}&0\\0&C_{22}\end{pmatrix}.\end{equation}
Previously, expressions were found for the Stroh matrices for an orthotropic bimaterial with a crack propagating at a constant speed, $v$, in \citet{Yang}, where the following parameters were defined
\[\kappa_{\gamma\beta}=\frac{C_{\gamma\beta}}{C_{66}}, \qquad \alpha_1=\sqrt{1-\frac{\rho v^2}{C_{11}}}, \qquad \alpha_2=\sqrt{1-\frac{\rho v^2}{C_{66}}},\]
\[\xi=\alpha_1 \alpha_2 \sqrt{\frac{\kappa_{11}}{\kappa_{22}}},\text{  and  } s=\frac{\alpha_2^2+\kappa_{11}\kappa_{22}\alpha_1^2-(1+\kappa_{12})^2}{2\alpha_1\alpha_2\sqrt{\kappa_{11}\kappa_{22}}}.\]
It is seen that the eigenvalues, with positive imaginary part, of equation \eqref{eigenstatic} are given by
\begin{equation}
p_{1,2}=\begin{cases}i\sqrt{\xi}\left(\sqrt{\frac{s+1}{2}}\pm\sqrt{\frac{s-1}{2}}\right),\quad\text{for }s\ge 1\\
\sqrt{\xi}\left(\pm\sqrt{\frac{1-s}{2}}+i\sqrt{\frac{1+s}{2}}\right), \quad\text{for } -1<s<1.\end{cases}
\end{equation}
Using the same normalisation as used in \citet{Yang} the matrices $\mathbf{A}$ and $\mathbf{L}$ are given by
\begin{equation}\label{matrixA}
\mathbf{A}=\begin{pmatrix}1&-\lambda_2^{-1}\\-\lambda_1&1\end{pmatrix},
\end{equation}\begin{equation}\label{matrixL}
\mathbf{L}=C_{66}\begin{pmatrix}p_1-\lambda_1&1-p_2\lambda_2^{-1}\\\kappa_{12}-\kappa_{22}p_1\lambda_1&\kappa_{22}p_2-\kappa_{12}\lambda_2^{-1}\end{pmatrix},
\end{equation}
where \[\lambda_\mu=\frac{\kappa_{11}\alpha_1^2+p_{\mu}^2}{(1+\kappa_{12})p_\mu}.\]

It is now possible to find an expression for the hermitian matrix $\mathbf{B}$
\begin{equation}\label{matrixB}
\mathbf{B}=i\mathbf{AL}^{-1}=\frac{1}{C_{66}R}\begin{pmatrix}\kappa_{22}\alpha_2^2\sqrt{2(1+s)/\xi}& i(\kappa_{22}-\kappa_{12}\alpha_2^2/\xi)\\
-i(\kappa_{22}-\kappa_{12}\alpha_2^2/\xi)&
\kappa_{22}\sqrt{2\xi(1+s)}\end{pmatrix},
\end{equation}
where $R$ is the generalized Rayleigh wave function given by \[R=\kappa_{22}(\kappa_{22}\xi-1+\alpha_2^2)-\kappa_{12}^2\alpha_2^2/\xi.\]
The Rayleigh wave speed of a material can be found by solving the equation, $R=0$.

The bimaterial matrix $\mathbf{H}$, from equation \eqref{hwdef}, has the form
\begin{equation}\label{matrixH}
\mathbf{H}=\begin{pmatrix}H_{11}&-i\beta\sqrt{H_{11}H_{22}}\\i\beta\sqrt{H_{11}H_{22}}&H_{22}\end{pmatrix}.
\end{equation}
From (\ref{matrixB}) it is seen that 
\[H_{11}=\left[\frac{\kappa_{22}\alpha_2^2\sqrt{2(1+s)/\xi}}{C_{66}R}\right]_I + \left[\frac{\kappa_{22}\alpha_2^2\sqrt{2(1+s)/\xi}}{C_{66}R}\right]_{II},\]
\[H_{22}=\left[\frac{\kappa_{22}\sqrt{2\xi(1+s)}}{C_{66}R}\right]_I + \left[\frac{\kappa_{22}\sqrt{2\xi(1+s)}}{C_{66}R}\right]_{II},\]
\[\beta\sqrt{H_{11}H_{22}}=\left[\frac{\kappa_{22}-\kappa_{12}\alpha_2^2/\xi}{C_{66}R}\right]_{II}- \left[\frac{\kappa_{22}-\kappa_{12}\alpha_2^2/\xi}{C_{66}R}\right]_I.\]
In order to compute the weight functions the eigenvalues and eigenvectors of (\ref{Heigen}) are required. Using the representation (\ref{matrixH}) it is found that
\begin{equation}\label{eigensol}
\mathbf{w}=\begin{pmatrix}-\frac{i}{2}\\\frac{1}{2}\sqrt{\frac{H_{11}}{H_{22}}}\end{pmatrix},\qquad \epsilon=\frac{1}{2\pi}\ln\left(\frac{1-\beta}{1+\beta}\right).
\end{equation}

Another key component for calculating the weight functions is the bimaterial matrix $\mathbf{W}$, defined in \eqref{hwdef}. Using (\ref{matrixB}) it is seen that
\begin{equation}\label{matrixW}
\mathbf{W}=\sqrt{H_{11}H_{22}} \begin{pmatrix}\delta_1\sqrt{\frac{H_{11}}{H_{22}}}&i{\gamma}\\
-i{\gamma}&\delta_2\sqrt{\frac{H_{22}}{H_{11}}}\end{pmatrix},
\end{equation}
where
\[\gamma=\frac{\left[\frac{\kappa_{22}-\kappa_{12}\alpha_2^2/\xi}{C_{66}R}\right]_I + \left[\frac{\kappa_{22}-\kappa_{12}\alpha_2^2/\xi}{C_{66}R}\right]_{II}} {\sqrt{H_{11}H_{22}}},\]
\[\delta_1=\frac{\left[\frac{\kappa_{22}\alpha_2^2\sqrt{2(1+s)/\xi}}{C_{66}R}\right]_I - \left[\frac{\kappa_{22}\alpha_2^2\sqrt{2(1+s)/\xi}}{C_{66}R}\right]_{II}}{H_{11}},\]
\[\delta_2=\frac{\left[\frac{\kappa_{22}\sqrt{2\xi(1+s)}}{C_{66}R}\right]_I - \left[\frac{\kappa_{22}\sqrt{2\xi(1+s)}}{C_{66}R}\right]_{II}}{H_{22}}.\]
\end{document}